\RequirePackage{ifpdf}
\ifpdf % We are running pdfTeX in pdf mode
\documentclass[pdftex]{sigma}
\else
\documentclass{sigma}
\fi

\newcommand{\rd}{\partial}

\newcommand{\Tr}{\operatorname{Tr}}

\newcommand{\CC}{\mathbf{C}}

\newcommand{\ZZ}{\mathbf{Z}}
\newcommand{\bspsi}{\boldsymbol{\psi}}
\newcommand{\Alpha}{\mathrm{A}}
\newcommand{\Beta}{\mathrm{B}}
\numberwithin{equation}{section}

\begin{document}

\allowdisplaybreaks

\renewcommand{\PaperNumber}{042}

\FirstPageHeading

\renewcommand{\thefootnote}{$\star$}

\ShortArticleName{Hamiltonian Structure of PI Hierarchy}
\ArticleName{Hamiltonian Structure of PI Hierarchy\footnote{This
paper is a contribution to the Vadim Kuznetsov Memorial Issue
`Integrable Systems and Related Topics'. The full collection is
available at
\href{http://www.emis.de/journals/SIGMA/kuznetsov.html}{http://www.emis.de/journals/SIGMA/kuznetsov.html}}}

\Author{Kanehisa TAKASAKI} \AuthorNameForHeading{K. Takasaki}
\Address{Graduate School of Human and Environmental Studies,
Kyoto University,\\
Yoshida, Sakyo, Kyoto 606-8501, Japan}
\Email{\href{mailto:takasaki@math.h.kyoto-u.ac.jp}{takasaki@math.h.kyoto-u.ac.jp}}

\ArticleDates{Received November 01, 2006, in f\/inal form February
13, 2007; Published online March 09, 2007}

\Abstract{The string equation of type $(2,2g+1)$ may be thought of
as a higher order analogue of the f\/irst Painlev\'e equation that
corresponds to the case of $g = 1$.  For $g > 1$, this equation is
accompanied with a f\/inite set of commuting isomonodromic
deformations, and they altogether form a hierarchy called the PI
hierarchy. This hierarchy gives an iso\-monod\-romic analogue of
the well known Mumford system. The Hamiltonian structure of the
Lax equations can be formulated by the same Poisson structure as
the Mumford system. A set of Darboux coordinates, which have been
used for the Mumford system, can be introduced in this hierarchy
as well.  The equations of motion in these Darboux coordinates
turn out to take a Hamiltonian form, but the Hamiltonians are
dif\/ferent from the Hamiltonians of the Lax equations (except for
the lowest one that corresponds to the string equation itself).}


\Classification{34M55; 35Q53; 37K20}

\section{Introduction}

The so called `string equations' were introduced in the discovery
of an exact solution of two-dimensional quantum gravity
\cite{BK90,DS90,GM90}.  Since it was obvious that these equations
are closely related to equations of the Painlev\'e and KdV type,
this breakthrough in string theory soon yielded a number of
studies from the point of view of integrable systems
\cite{Douglas90,Moore90,FKN91,KS91,Schwarz91}.

The string equations are classif\/ied by a pair $(q,p)$ of coprime
positive integers. The simplest case of $(q,p) = (2,3)$ is nothing
but the f\/irst Painlev\'e equation
\begin{gather*}
  \frac{1}{4}u_{xx} + \frac{3}{4}u^2 + x = 0,
\end{gather*}
and the equations of type $(2,p)$ for $p = 5,7,\ldots$ may be
thought of as higher order analogues thereof. Unlike the case of
type $(2,3)$, these higher order PI equations are accompanied with
a f\/inite number of commuting f\/lows, which altogether form a
kind of f\/inite-dimensional `hierarchy'.  This hierarchy is
referred to as `the PI hierarchy' in this paper. (`PI' stands for
the f\/irst Painlev\'e equation).

The PI hierarchy can be characterized as a reduction of the KdV
(or KP) hierarchy. This is also the case for the string equations
of all types. The role of the string equation in this reduction
resembles that of the equation of commuting pair of dif\/ferential
operators \cite{BC22-31}.  It is well known that the equation of
commuting pairs, also called `the stationary Lax equation',
characterizes algebro-geometric solutions of the KP hierarchy
\cite{DMN76,Krichever77}.  The reduction by the string equation,
however, is drastically dif\/ferent in its nature. Namely, whereas
the commuting pair equation imposes translational symmetries to
the KP hierarchy, the string equation is related to Virasoro (and
even larger $W_{1+\infty}$) symmetries \cite{FKN91}. In the Lax
formalism of the KP hierarchy \cite{SS82,DJKM83,SW85}, the latter
symmetries are realized by the Orlov--Schulman operator
\cite{OS86}, which turns out to be an extremely useful tool for
formulating the string equation and the accompanied commuting
f\/lows \cite{AvM92}.

The string equation and the accompanied commuting f\/lows may be
viewed as a system of isomonodromic deformations.  This is
achieved by reformulating the equations as Lax equations of a
polynomial $L$-matrix (a $2 \times 2$ matrix in the case of the PI
hierarchy). This Lax formalism may be compared with the Lax
formalism of the well known Mumford system \cite{Mumford84}. Both
systems have substantially the same $2 \times 2$ matrix
$L$-matrix, which is denoted by $V(\lambda)$ in this paper.
$\lambda$~is a spectral parameter on which $V(\lambda)$ depends
polynomially. The dif\/ference of these systems lies in the
structure of the Lax equations. The Lax equations of the Mumford
system take such a~form as
\begin{gather*}
  \rd_tV(\lambda) = [U(\lambda),V(\lambda)],
\end{gather*}
where $U(\lambda)$ is also a $2 \times 2$ matrix of polynomials in
$\lambda$.  (Note that we show just one of the Lax equations
representatively.)  Obviously, this is an isospectral system.  On
the other hand, the corresponding Lax equations of the PI
hierarchy have an extra term on the right hand side:
\begin{gather*}
  \rd_tV(\lambda)
  = [U(\lambda),V(\lambda)] + U'(\lambda), \qquad
  U'(\lambda) = \rd_\lambda U(\lambda).
\end{gather*}
The extra term $U'(\lambda)$ breaks isospectrality. This is
actually a common feature of Lax equations that describe
isomonodromic deformations.

We are concerned with the Hamiltonian structure of this kind of
isomonodromic systems. As it turns in this paper, the PI hierarchy
exhibits some new aspects of this issue. Let us brief\/ly show an
outline.

\looseness=-1 The Hamiltonian structure of the Mumford system is
more or less well known \cite{Beauville90,NS01}. The Poisson
brackets of the matrix elements of the $L$-matrix take the form of
`generalized linear brac\-kets' \cite{EEKT94}. (Actually, this
system has a multi-Hamiltonian structure \cite{MM91,FMPZ00}, but
this is beyond the scope of this paper.)  The Lax equations can be
thereby expressed in the Hamiltonian form
\begin{gather*}
  \rd_tV(\lambda) = \{V(\lambda),H\}.
\end{gather*}

Since the Lax equations of the PI hierarchy have substantially the
same $L$-matrix as the Mumford system, we can borrow its Poisson
structure.  In fact,  the role of the Poisson structure is simply
to give an identity of the form
\begin{gather*}
  [U(\lambda),V(\lambda)] = \{V(\lambda),H\}.
\end{gather*}
We can thus rewrite the Lax equations as
\begin{gather*}
  \rd_tV(\lambda)
  = \{V(\lambda),H\} + U'(\lambda),
\end{gather*}
leaving the extra term intact. This is a usual understanding of
the Hamiltonian structure of isomonodromic systems such as the
Schlesinger system (see, e.g., \cite[Appendix 5]{JMMS80}). This
naive prescription, however, leads to a dif\/f\/iculty when we
consider a set of Darboux coordinates called `spectral Darboux
coordinates' and attempt to rewrite the Lax equations to a
Hamiltonian system in these coordinates.

The notion of spectral Darboux coordinates originates in the
pioneering work of Flaschka and McLaughlin \cite{FM76}, which was
later reformulated by Novikov and Veselov \cite{NV84} in a more
general form.  As regards the Mumford system, this notion lies in
the heart of the classical algebro-geometric approach
\cite{Mumford84}. The Montreal group \cite{AHH93} applied these
coordinates to separation of variables of various isospectral
systems with a rational $L$-matrix.  Their idea was generalized by
Sklyanin \cite{Sklyanin95} to a wide range of integrable systems
including quantum integrable systems. On the other hand, spectral
Darboux coordinates were also applied to isomonodromic systems
\cite{Harnad94,HW96,Takasaki03,DM03}.

We can consider spectral Darboux coordinates for the PI hierarchy
in exactly the same way as the case of the Mumford system. It will
be then natural to attempt to derive equations of motions in those
Darboux coordinates. Naive expectation will be that those
equations of motion become a Hamiltonian system with the same
Hamiltonian $H$ as the Lax equation. This, however, turns out to
be wrong (except for the lowest part of the hierarchy, namely, the
string equation itself). The fact is that the extra term
$U'(\lambda)$ in the Lax equation gives rise to extra terms in the
equations of motions in the Darboux coordinates. Thus, not only
the naive expectation is negated, it is also not evident whether
those equations of motion take a Hamiltonian form with a suitable
Hamiltonian. This is the aforementioned dif\/f\/iculty.

The goal of this paper is to show that those equations of motion
are indeed a Hamiltonian system.  As it turns out, the correct
Hamiltonian $K$ can be obtained by adding a correction term
$\Delta H$ to $H$ as
\begin{gather*}
  K = H + \Delta H.
\end{gather*}
This is a main conclusion of our results. It is interesting that
the Hamiltonian of the lowest f\/low of the hierarchy (namely, the
string equation itself) is free from the correction term.

\looseness=-1 Let us mention that this kind of correction terms
take place in some other isomonodromic systems as well. An example
is the Garnier system (so named and) studied by Okamoto
\cite{Okamoto86}. The Garnier system is a multi-dimensional
generalization of the Painlev\'e equations (in particular, the
sixth Painlev\'e equation), and has two dif\/ferent
interpretations as isomonodromic deformations.  One is based on a
second order Fuchsian equation. Another interpretation is the $2
\times 2$ Schlesinger system, from which the Garnier system can be
derived as a Hamiltonian system for a special set of Darboux
coordinates.  It is easy to see that these Darboux coordinates are
nothing but spectral Darboux coordinates in the aforementioned
sense \cite{Harnad94,HW96}, and that the Hamiltonians are the
Hamiltonians of the Schlesinger system \cite[Appendix 5]{JMMS80}
plus correction terms.  These observations on the Garnier system
have been generalized by Dubrovin and Mazzocco \cite{DM03} to the
Schlesinger system of an arbitrary size. A similar structure of
Hamiltonians can be found in a~`degenerate' version of the Garnier
system studied by Kimura \cite{Kimura89} and Shimomura
\cite{Shimomura00}.  Actually, this system coincides with the PI
hierarchy associated with the string equation of type $(2,5)$.

This paper is organized as follows. In Section 2, we introduce the
string equations of type $(2,p)$, and explain why they can be
viewed as a higher order analogues of the f\/irst Painlev\'e
equations. In Section 3, these equations are cast into a $2 \times
2$ matrix Lax equation. Section 4 is a brief review of the KdV and
KP hierarchies. In Section 5, the PI hierarchy is formulated as a
reduction of the KP (or KdV) hierarchy, and converted to $2 \times
2$ matrix Lax equations. In Section 6, we introduce the notion of
spectral curve and consider the structure of its def\/ining
equation  in detail. Though the results of this section appear to
be rather technical, they are crucial to the description of
Hamiltonians in spectral Darboux coordinates. Section 7 deals with
the Hamiltonian structure of the Lax equations.  In Section 8, we
introduce spectral Darboux coordinates, and in Section 9, identify
the Hamiltonians in these coordinates.  In Section 10, these
results are illustrated for the f\/irst three cases of $(q,p) =
(2,3), (2,5),(2,7)$.

\section{String equation as higher order PI equation}

Let $q$ and $p$ be a pair of coprime positive integers.  The
string equation of type $(q,p)$ takes the commutator form
\cite{Douglas90}
\begin{gather}
  [Q,P] = 1
\label{[Q,P]=1}
\end{gather}
for a pair of ordinary dif\/ferential operators
\begin{gather*}
  Q = \rd_x^q + g_2\rd_x^{q-2} + \cdots + g_q, \qquad
  P = \rd_x^p + f_2\rd_x^{p-2} + \cdots + f_p
\end{gather*}
of order $q$ and $p$ in one-dimensional spatial variable $x$
($\rd_x = \rd/\rd x$). In the following, we consider the equation
of type $(q,p) = (2,2g+1)$, $g = 1,2,\ldots$. We shall see that
$g$ is equal to the genus of an underlying algebraic curve
(spectral curve).

The simplest case, i.e., $(q,p) = (2,3)$, consists of operators of
the form
\begin{gather*}
  Q = \rd_x^2 + u, \qquad
  P = \rd_x^3 + \frac{3}{2}u\rd_x + \frac{3}{4}u_x,
\end{gather*}
where the subscript means a derivative as
\begin{gather*}
  u_x = \frac{\rd u}{\rd x},\qquad
  u_{xx} = \frac{\rd^2 u}{\rd x^2}, \quad
  \ldots.
\end{gather*}
The string equation \eqref{[Q,P]=1} for these operators reduces to
the third-order equation
\begin{gather*}
  \frac{1}{4}u_{xxx} + \frac{3}{2}uu_x + 1 = 0.
%\label{PI-eqx}
\end{gather*}
We can integrate it once, eliminating the integration constant by
shifting $x \to x + \text{const}$, and obtain the f\/irst
Painlev\'e equation
\begin{gather}
  \frac{1}{4}u_{xx} + \frac{3}{4}u^2 + x = 0.
\label{PI-eq}
\end{gather}

The setup for the general case of type $(2,2g+1)$ relies on the
techniques originally developed for the KdV hierarchy and its
generalization \cite{GD75-77,Manin79,Adler79,Wilson79}. The basic
tools are the fractional powers
\begin{gather*}
  Q^{n+1/2}  = \rd_x^{2n+1} + \frac{2n+1}{2}u\rd_x^{2n-1}
    + \cdots + R_{n+1}\rd_x^{-1} + \cdots
\end{gather*}
of $Q = \rd_x^2 + u$.  The fractional powers are realized as
pseudo-dif\/ferential operators. The coef\/f\/i\-cient~$R_{n+1}$
of $\rd_x^{-1}$ is a dif\/ferential polynomials of $u$ called the
Gelfand--Dickey polynomial:
\begin{gather*}
R_0 = 1, \qquad
  R_1 = \frac{u}{2}, \qquad
  R_2 = \frac{1}{8}u_{xx} + \frac{3}{8}u^2, \nonumber
  \\
R_3 = \frac{1}{32}u_{xxxx} + \frac{3}{16}uu_{xx}
        + \frac{5}{32}u_x^2 + \frac{5}{16}u^3,  \quad \ldots.
%\label{GD-poly}
\end{gather*}
For all $n$'s, the highest order term in $R_n$ is linear and
proportional to $u^{(2n-2)}$:
\begin{gather*}
  R_n = \frac{1}{2^{2n+2}}u^{(2n-2)} + \cdots.
\end{gather*}

As in the construction of the KdV hierarchy, we introduce the
dif\/ferential operators
\begin{gather*}
  B_{2n+1} = \bigl(Q^{n+1/2}\bigr)_{\ge 0},
  \qquad n = 0,1,\ldots.
\end{gather*}
where $(\ )_{\ge 0}$ stands for the projection of a
pseudo-dif\/ferential operator to a dif\/ferential operator.
Similarly, we use the notation $(\ )_{<0}$ for the projection to
the part of negative powers of $\rd_x$, i.e.,
\begin{gather*}
  \Biggl(\sum_{j\in\ZZ} a_j\rd_x^j\Biggr)_{\ge 0}  = \sum_{j\ge 0} a_j\rd_x^j, \qquad
  \Biggl(\sum_{j\in\ZZ} a_j\rd_x^j\Biggr)_{<0}  = \sum_{j<0} a_j\rd_x^j.
\end{gather*}
These dif\/ferential operators have the special property that the
commutator with $Q$ is of order zero. More precisely, we have the
identity
\begin{gather}
  [B_{2n+1},Q] = 2R_{n+1,x}.
\label{[B,Q]}
\end{gather}
Consequently, if we choose $P$ to be a linear combination of these
operators as
\begin{gather}
  P = B_{2g+1} + c_1B_{2g-1} + \cdots + c_gB_1
\label{P=B+...}
\end{gather}
with constant coef\/f\/icients $c_1,\ldots,c_g$, the commutator
with $Q$ reads
\begin{gather*}
  [Q,P] = - 2R_{g+1,x} - 2c_1R_{g,x} - \cdots - 2c_gR_{1,x}.
\end{gather*}
The string equation \eqref{[Q,P]=1} thus reduces to
\begin{gather}
  2R_{g+1,x} + 2c_1R_{g,x} + \cdots + 2c_gR_{1,x} + 1 = 0.
\label{higher-PI-eqx}
\end{gather}
This equation can be integrated to become the equation
\begin{gather}
  2R_{g+1} + 2c_1R_g + \cdots + 2c_gR_1 + x = 0,
\label{higher-PI-eq}
\end{gather}
which gives a higher order analogue (of order $2g$) of the f\/irst
Painlev\'e equation \eqref{PI-eq}.

When we consider a hierarchy of commuting f\/lows that preserve
the string equation, the coef\/f\/icients $c_1,\ldots,c_g$ play
the role of time variables.  For the moment, they are treated as
constants.

\section{Matrix Lax formalism  of string equation}

The string equation \eqref{[Q,P]=1} is accompanied with the
auxiliary linear problem
\begin{gather}
  Q\psi = \lambda\psi, \qquad
  P\psi = \rd_\lambda\psi
\label{QP-lineq}
\end{gather}
with a spectral parameter $\lambda$ ($\rd_\lambda =
\rd/\rd\lambda$). We can rewrite this linear problem to a $2
\times 2$ matrix form \cite{Moore90} using various properties of
the operators $B_{2n+1}$ and the Gelfand--Dickey polynomials~$R_n$
\cite{GD75-77,Manin79,Adler79,Wilson79} as follows.

The f\/irst equation of \eqref{QP-lineq} can be readily converted
to the matrix form
\begin{gather}
  \rd_x\bspsi = U_0(\lambda)\bspsi,
\label{bspsi-x-lineq}
\end{gather}
where
\begin{gather*}
  \bspsi
  = \left(\begin{matrix}
    \psi \\
    \psi_x
    \end{matrix}\right),
  \qquad
  U_0(\lambda)
  = \left(\begin{matrix}
    0 & 1 \\
    \lambda - u & 0
    \end{matrix}\right).
\end{gather*}

To rewrite the second equation of \eqref{QP-lineq}, we use the
`$Q$-adic' expansion formula
\begin{gather}
  B_{2n+1}
  = \sum_{m=0}^n \biggl(R_m\rd_x - \frac{1}{2}R_{m,x}\biggr)Q^{n-m}
\label{B-expansion}
\end{gather}
of $B_{2n+1}$'s.  By this formula, $B_{2n+1}\psi$ becomes a linear
combination of $\psi$ and $\psi_x$ as
\begin{gather*}
  B_{2n+1}\psi
  = \sum_{m=0}^n\biggl(R_m\psi_x - \frac{1}{2}R_{m,x}\psi\biggr)
  = R_m(\lambda)\psi_x - \frac{1}{2}R_m(\lambda)_x\psi,
\end{gather*}
where $R_n(\lambda)$ stand for the auxiliary polynomials
\begin{gather}
  R_n(\lambda)
  = \lambda^n + R_1\lambda^{n-1} + \cdots + R_n,
  \qquad n = 0,1,\ldots,
\label{R_n(l)-def}
\end{gather}
that play a central role throughout this paper. Since $P$ is a
linear combination of $B_{2n+1}$'s as~\eqref{P=B+...} shows, we
can express $P\psi$ as
\begin{gather*}
  P\psi = \alpha(\lambda)\psi + \beta(\lambda)\psi_x
\end{gather*}
with the coef\/f\/icients
\begin{gather}
  \beta(\lambda) = R_g(\lambda) + c_1R_{g-1}(\lambda)
    + \cdots + c_gR_0(\lambda), \qquad
  \alpha(\lambda) = - \frac{1}{2}\beta(\lambda)_x.
\label{beta-alpha-def}
\end{gather}
Moreover, dif\/ferentiating this equation by $x$ and substituting
$\psi_{xx} = (\lambda - u)\psi$, we can express~$P\psi_x$ as
\begin{gather*}
  P\psi_x
  = (P\psi)_x
  = \gamma(\lambda)\psi - \alpha(\lambda)_x\psi_x,
\end{gather*}
where
\begin{gather}
  \gamma(\lambda)
  = \alpha(\lambda)_x + (\lambda - u)\beta(\lambda)
  = - \frac{1}{2}\beta(\lambda)_{xx}
    + (\lambda - u)\beta(\lambda).
\label{gamma-def}
\end{gather}
The second equation of the auxiliary linear problem
\eqref{QP-lineq} can be thus converted to the matrix form
\begin{gather}
  \rd_\lambda\bspsi = V(\lambda)\bspsi
\label{bspsi-l-lineq}
\end{gather}
with the coef\/f\/icient matrix
\begin{gather*}
  V(\lambda)
  = \left(\begin{matrix}
    \alpha(\lambda) & \beta(\lambda) \\
    \gamma(\lambda) & - \alpha(\lambda)
    \end{matrix}\right).
\end{gather*}

The dif\/ferential equation \eqref{bspsi-x-lineq} in $x$ may be
thought of as def\/ining isomonodromic deformations of the matrix
ODE \eqref{bspsi-l-lineq} in $\lambda$ with polynomial
coef\/f\/icients. The associated Lax equation reads
\begin{gather}
  \rd_xV(\lambda)
  = [U_0(\lambda),V(\lambda)] + U_0'(\lambda),
\label{V-x-Laxeq}
\end{gather}
where the last term stands for the $\lambda$-derivative of
$U_0(\lambda)$, i.e.,
\begin{gather*}
  U_0'(\lambda)
  = \rd_\lambda U_0(\lambda)
  = \left(\begin{matrix}
    0 & 0 \\
    1 & 0
    \end{matrix}\right).
\end{gather*}
The presence of such an extra term other than a matrix commutator
is a common characteristic of Lax equations for isomonodromic
deformations in general.

In components, the Lax equation consists of the three equations
\begin{gather*}
\rd_x\alpha(\lambda)  = \gamma(\lambda) - (\lambda -
u)\beta(\lambda), \qquad \rd_x\beta(\lambda) = -
2\alpha(\lambda),\nonumber \qquad \rd_x\gamma(\lambda) = 2(\lambda
- u)\alpha(\lambda) + 1.
%\label{V-x-Laxeq2}
\end{gather*}
The f\/irst and second equations can be solved for
$\alpha(\lambda)$ and $\gamma(\lambda)$; the outcome is just the
def\/inition of these polynomials in \eqref{beta-alpha-def} and
\eqref{gamma-def}.  Upon eliminating $\alpha(\lambda)$ and
$\gamma(\lambda)$ by these equations, the third equation turns
into the equation{\samepage
\begin{gather}
  \frac{1}{2}\beta(\lambda)_{xxx}
  - 2(\lambda- u)\beta(\lambda)_x
  + u_x\beta(\lambda) + 1
  = 0
\label{beta-x-eq}
\end{gather}
for $\beta(\lambda)$ only.}

Let us examine \eqref{beta-x-eq} in more detail.  By the
def\/inition in \eqref{beta-alpha-def}, $\beta(\lambda)$ is a
polynomial of the form
\begin{gather*}
  \beta(\lambda)
  = \lambda^g + \beta_1\lambda^{g-1}
    + \cdots + \beta_g
\end{gather*}
with the coef\/f\/icients
\begin{gather}
  \beta_n = R_n + c_1R_{n-1} + \cdots + c_nR_0,
  \qquad n = 1,\ldots,g.
\label{beta_n=R_n+...}
\end{gather}
Upon expanded into powers of $\lambda$, \eqref{beta-x-eq} yield
the equations
\begin{gather*}
  \frac{1}{2}\beta_{n,xxx} + 2u\beta_{n,x}
  + u_x\beta_n - 2\beta_{n+1,x} = 0,
  \qquad n = 1,\ldots,g-1,
\end{gather*}
and
\begin{gather*}
  \frac{1}{2}\beta_{g,xxx} + 2u\beta_{g,x}
  + u_x\beta_g + 1
  = 0.
\end{gather*}
Recalling here the well known Lenard recursion formula
\begin{gather}
  R_{x,n+1}
  = \frac{1}{4}R_{n,xxx} + uR_{n,x}
    + \frac{1}{2}u_xR_n,
\label{Lenard}
\end{gather}
one can see that the forgoing equations for $n = 1,\ldots,g-1$ are
identities, and that the latter one is equivalent to
\eqref{higher-PI-eqx}.

\section{KdV and KP hierarchies}

We shall derive the PI hierarchy from the KdV or KP hierarchy.
Let us brief\/ly review some basic stuf\/f of these hierarchies
\cite{SS82,DJKM83,SW85}. Of particular importance is the notion of
the Orlov--Schulman operator \cite{OS86} that supplements the
usual Lax formalism of these hierarchies.

\subsection{KdV hierarchy from KP hierarchy}

Let $L$ denote the Lax operator
\begin{gather*}
  L = \rd_x + u_2\rd_x^{-1} + u_3\rd_x^{-2} + \cdots
\end{gather*}
of the KP hierarchy.  $L$ obeys the Lax equations
\begin{gather*}
  \rd_n L = [B_n,L], \qquad
  B_n = (L^n)_{\ge 0},
\end{gather*}
in an inf\/inite number of time variables $t_1,t_2,\ldots$ ($\rd_n
= \rd/\rd t_n$). As usual, we identify $t_1$ with $x$.

The KdV hierarchy can be derived from the KP hierarchy by imposing
the constraint
\begin{gather*}
  (L^2)_{<0} = 0.
\end{gather*}
Under this constraint, $Q = L^2$ becomes a dif\/ferential operator
of the form
\begin{gather*}
  Q = \rd_x^2 + u, \qquad u = 2u_2,
\end{gather*}
all even f\/lows are trivial in the sense that
\begin{gather*}
  \rd_{2n}L = [L^{2n},L] = 0,
\end{gather*}
and the whole hierarchy reduces to the Lax equations
\begin{gather}
  \rd_{2n+1}Q = [B_{2n+1},Q], \qquad
  B_{2n+1} = (Q^{n+1/2})_{\ge 0},
\end{gather}
of the KdV hierarchy.   By \eqref{[B,Q]}, these Lax equations
further reduce to the evolution equations
\begin{gather}
  \rd_{2n+1}u = 2R_{n+1,x}
\label{u-t-eq}
\end{gather}
for $u$.

\subsection[Orlov-Schulman operator]{Orlov--Schulman operator}

The Orlov--Schulman operator is a pseudo-dif\/ferential operator
(of inf\/inite order) of the form
\begin{gather*}
  M = \sum_{n=2}^\infty nt_nL^{n-1} + x
      + \sum_{n=1}^\infty v_nL^{-n-1},
\end{gather*}
that obeys the Lax equations
\begin{gather}
  \rd_nM = [B_n,M]
\label{M-t-Laxeq}
\end{gather}
and the commutation relation
\begin{gather}
  [L,M] = 1.
\label{[L,M]=1}
\end{gather}

The existence of such an operator can be explained in the language
of the auxiliary linear system
\begin{gather}
  L\psi = z\psi, \qquad
  \rd_n\psi = B_n\psi,\;
\label{KP-LB-lineq}
\end{gather}
of the KP hierarchy.  This linear system has a (formal) solution
of the form
\begin{gather}
  \psi
  = \Biggl(1 + \sum_{j=1}^\infty w_jz^{-j}\Biggr)
    \exp\Biggl(xz + \sum_{n=2}^\infty t_nz^n\Biggr),
\label{normal-psi}
\end{gather}
One can rewrite this solution as
\begin{gather*}
  \psi = W\exp\Biggl(xz + \sum_{n=2}^\infty t_nz^n\Biggr),
\end{gather*}
where $W$ is a pseudo-dif\/ferential operator (called the
`dressing operator' or the Sato--Wilson operator) of the form
\begin{gather*}
  W = 1 + \sum_{j=1}^\infty w_j\rd_x^{-j}
\end{gather*}
that satisf\/ies the evolution equations
\begin{gather}
  \rd_nW = - (W\cdot\rd_x^n\cdot W^{-1})_{<0}W
\label{KP-Sato-eq}
\end{gather}
and the algebraic relations
\begin{gather*}
  L = W\cdot\rd_x\cdot W^{-1},\qquad
  B_n = (W\cdot\rd_x^{n}\cdot W^{-1})_{\ge 0}.
\end{gather*}
One can now def\/ine $M$ as
\begin{gather*}
  M = W\left(\sum_{n=2}^\infty nt_n\rd_x^{n-1}
             + x\right)W^{-1},
\end{gather*}
which turns out to satisfy the foregoing equations
\eqref{M-t-Laxeq}, \eqref{[L,M]=1} and the auxiliary linear
equation
\begin{gather}
  M\psi = \rd_z\psi.
\label{KP-M-lineq}
\end{gather}

\begin{remark}
In the reduction to the KdV hierarchy, the $\rd_x^{-1}$-part of
\eqref{KP-Sato-eq} gives the equation
\begin{gather}
  \rd_{2n+1}w_1 = - R_{n+1}.
\label{w_1-t-eq}
\end{gather}
Since $u = - 2w_{1,x}$, this equation may be thought of as a
once-integrated form of \eqref{u-t-eq}.
\end{remark}

\section{PI hierarchy}

We now formulate the PI hierarchy as a reduction of the KP
hierarchy, and rewrite it to a $2 \times 2$ matrix Lax equation,
as we have done for the string equation itself. The string
equation and the commuting f\/lows of the PI hierarchy are thus
unif\/ied to a system of multi-time isomonodromic deformations.

\subsection{PI hierarchy from KP hierarchy}

To derive the string equation from the KP hierarchy, we impose the
constraints \cite{AvM92}
\begin{gather}
  (Q)_{<0} = 0, \qquad
  (P)_{<0} = 0
\label{PI-constraint}
\end{gather}
on the operators
\begin{gather*}
  Q = L^2, \qquad
  P = \frac{1}{2}ML^{-1}
    = \sum_{n=1}^\infty nt_nL^{n-2}
      + \sum_{n=1}^\infty v_nL^{-n-2}.
\end{gather*}
This leads to the following consequences.

\begin{enumerate}\itemsep=0pt
\item As it follows from the commutation relation \eqref{[L,M]=1}
of $L$ and $M$, these operators obey the commutation relation
$[Q,P] = 1$. \item Under the f\/irst constraint of
\eqref{PI-constraint}, $Q$ becomes the Lax operator $\rd_x^2 + u$
of the KdV hierarchy.  In the following, as usual, we suppress the
time variables with even indices, i.e.,
\begin{gather*}
  t_2 = t_4 = \cdots = 0.
\end{gather*}
\item The second constraint of \eqref{PI-constraint} implies that
$P$ is a dif\/ferential operator (of inf\/inite order) of the form
\begin{gather*}
  P = \frac{1}{2}\bigl(ML^{-1}\bigr)_{\ge 0}
    = \sum_{n=1}^\infty
      \frac{2n+1}{2}t_{2n+1}B_{2n+1}.
\end{gather*}
Therefore, if we set
\begin{gather}
  t_{2g+3} = \frac{2}{2g+1}, \;
  t_{2g+5} = t_{2g+7} = \cdots = 0,
\label{higher-t}
\end{gather}
we are left with a dif\/ferential operator of the form
\eqref{P=B+...} with the coef\/f\/icients $c_1,\ldots,c_g$
depending on the time variables as
\begin{gather}
  c_n = c_n(t) = \frac{2n+1}{2}t_{2n+1},
  \qquad n = 1,\ldots,g.
\label{c_n(t)}
\end{gather}
\item The auxiliary linear equations \eqref{KP-LB-lineq} and
\eqref{KP-M-lineq} imply the linear equations
\begin{gather*}
  Q\psi = z^2\psi, \qquad
  P\psi = \frac{1}{2}z^{-1}\rd_z\psi.
\end{gather*}
These linear equations can be identif\/ied with the auxiliary
linear equations \eqref{QP-lineq} of the string equation if we
def\/ine $\lambda$ as
\begin{gather*}
  \lambda = z^2.
\end{gather*}
\end{enumerate}

We can thus recover, from the KP hierarchy, the string equation
\eqref{[Q,P]=1} of type $(2,2g+1)$ along with $g$ extra commuting
f\/lows
\begin{gather*}
  \rd_{2n+1}Q = [B_{2n+1},Q], \qquad
  \rd_{2n+1}P = [B_{2n+1},P], \qquad
  n = 1,\ldots,g.
%\label{PI-hierarchy}
\end{gather*}
We call this system the PI hierarchy. Let us mention that this
hierarchy was f\/irst discovered in a more direct way
\cite{Douglas90}.  Compared with that approach, the approach from
the KP hierarchy \cite{AvM92} is more transparent.

\begin{remark}
This hierarchy thus contains $g + 1$ f\/lows with time variables
$t_1 (= x),\,t_3,\,\ldots,\, t_{2g+1}$. The last f\/low in
$t_{2g+1}$, however, turns out to be spurious, i.e., can be
absorbed by other f\/lows (see below).  We shall suppress this
f\/low when we consider the Hamiltonian structure.
\end{remark}

\subsection[Matrix Lax formalism of commuting flows]{Matrix Lax formalism of commuting f\/lows}

We can again use the $Q$-adic expansion formula
\eqref{B-expansion} of $B_{2n+1}$ to rewrite the auxiliary linear
equations
\begin{gather*}
  \rd_{2n+1}\psi = B_{2n+1}\psi,
  \qquad n = 1,\ldots,g,
\end{gather*}
of the PI hierarchy to linear equations
\begin{gather}
  \rd_{2n+1}\bspsi  = U_n(\lambda)\bspsi
\label{bspsi-t-lineq}
\end{gather}
for $\bspsi$.  The matrix elements of the coef\/f\/icient matrix
\begin{gather*}
  U_n(\lambda)
  = \left(\begin{matrix}
    \Alpha_n(\lambda) & \Beta_n(\lambda) \\
    \Gamma_n(\lambda) & - \Alpha_n(\lambda)
    \end{matrix}\right)
\end{gather*}
are the following polynomials in $\lambda$:
\begin{gather}
\Beta_n(\lambda) = R_n(\lambda), \qquad \Alpha_n(\lambda) = -
\frac{1}{2}R_n(\lambda)_x,\nonumber
\\
\Gamma_n(\lambda)
  = - \frac{1}{2}R_n(\lambda)_{xx} + (\lambda - u)R_n(\lambda).
\label{U_n-def}
\end{gather}
\eqref{bspsi-x-lineq} can be included in these linear equations as
a special case with $n = 0$ ($t_1 = x$).

As one can see by comparing these matrix elements with those of
$U_n(\lambda)$'s def\/ined in \eqref{beta-alpha-def} and
\eqref{gamma-def}, $V(\lambda)$ is a linear combination of
$U_n(\lambda)$'s:
\begin{gather}
  V(\lambda)
  = U_g(\lambda) + c_1(t)U_{g-1}(\lambda)
    + \cdots + c_g(t)U_0(\lambda).
\label{V=U+...}
\end{gather}
We shall consider implications of this linear relation later on.

Having obtained the full set of auxiliary linear equations
\eqref{bspsi-x-lineq}, \eqref{bspsi-l-lineq},
\eqref{bspsi-t-lineq} in a $2 \times 2$ matrix form, we can now
reformulate the commuting f\/lows of the PI hierarchy as the
isomonodromic matrix Lax equations
\begin{gather}
  \rd_{2n+1}V(\lambda)
  = [U_n(\lambda),V(\lambda)] + U_n'(\lambda),
  \qquad n = 0,1,\ldots,g,
\label{V-t-Laxeq}
\end{gather}
including \eqref{V-x-Laxeq} as the case with $n = 0$ ($t_1 = x$).
Note that the Lax equations of the higher f\/lows, too, have an
extra term
\begin{gather*}
  U_n'(\lambda) = \rd_\lambda U_n(\lambda)
\end{gather*}
other than the matrix commutator $[U_n(\lambda),V(\lambda)]$.

\subsection{Variant of self-similarity}

Nowadays it is widely known that many isomonodromic equations can
be derived from soliton equations as `self-similar reduction'
\cite{ARS80,FN80}. The transition from the KP hierarchy to the PI
hierarchy is actually a variant of self-similar reduction.

Recall that $P$ is a linear combination of $B_n$'s as
\eqref{P=B+...} shows. This implies that the left hand side of the
auxiliary linear equation $P\psi = \rd_\lambda\psi$ can be
expressed as
\begin{gather*}
  P\psi = (\rd_{2g+1} + c_1(t)\rd_{2g-1}
           + \cdots + c_g(t)\rd_{1})\psi.
\end{gather*}
The auxiliary linear equation thus turns into a kind of linear
constraint:
\begin{gather}
  (\rd_{2g+1} + c_1(t)\rd_{2g-1} + \cdots + c_g(t)\rd_{1})\psi
  = \rd_\lambda\psi.
\label{psi-constraint}
\end{gather}

If we were considering the equation
\begin{gather*}
  [Q,P] = 0
%\label{[Q,P]=0}
\end{gather*}
of commuting pair of dif\/ferential operators
\cite{BC22-31,DMN76,Krichever77}, all $c_n$'s would be constants,
and \eqref{psi-constraint} would mean the existence of stationary
directions (in other words, translational symmetries) in the whole
time evolutions of the KdV hierarchy. It is well known that this
condition characterizes algebro-geometric solutions of the KdV
hierarchy.

In the present setting, where $c_n$'s are not constant but
variables as \eqref{c_n(t)} shows, \eqref{psi-constraint} may be
thought of as a variant of self-similarity condition. We can see
some more manifestation of this condition.  For instance,
$V(\lambda)$ satisf\/ies the linear equation
\begin{gather*}
  (\rd_{2g+1} + c_1(t)\rd_{2g-1}
  + \cdots + c_g(t)\rd_1)V(\lambda)
  = V'(\lambda)
%\label{V-constraint}
\end{gather*}
as one can deduce from the Lax equations \eqref{V-t-Laxeq} and the
linear relation \eqref{V=U+...} among their coef\/f\/icients.
Similarly, $u$ satisf\/ies a similar equation
\begin{gather*}
  (\rd_{2g+1} + c_1(t)\rd_{2g-1} + \cdots + c_g(t)\rd_1)u = 1
%\label{u-constraint}
\end{gather*}
as a consequence of \eqref{higher-PI-eqx} and \eqref{u-t-eq}.
These equations show, in particular, that the $t_{g+1}$-f\/low is
spurious.

\begin{remark}
\eqref{psi-constraint} stems from the Virasoro constraints
associated with the string equation. The Virasoro constraints are
usually formulated in the language of the $\tau$-function
\cite{FKN91}. Reformulated in terms of $\psi$, the lowest one (the
$L_{-1}$-constraint) of those constraints becomes the linear
equation
\begin{gather*}
  \sum_{n=0}^\infty
  \frac{2n+1}{2}t_{2n+1}\rd_{2n-1}\psi
  = \rd_\lambda\psi.
\end{gather*}
This turns into the aforementioned constraint
\eqref{psi-constraint} if the higher time variables
$t_{2g+3},t_{2g+5},\ldots$ are set to the special values of
\eqref{higher-t}.
\end{remark}

\section{Building blocks of spectral curve}

\subsection{Equation of spectral curve}

We now consider the spectral curve of the matrix $V(\lambda)$.
This curve is def\/ined by the characte\-ristic equation
\begin{gather*}
  \det(\mu I - V(\lambda))
  = \mu^2 + \det V(\lambda) = 0
\end{gather*}
or, more explicitly, by the equation
\begin{gather*}
  \mu^2 = h(\lambda)
  = \alpha(\lambda)^2 + \beta(\lambda)\gamma(\lambda).
\end{gather*}
Since $\alpha(\lambda),\beta(\lambda), \gamma(\lambda)$ are
polynomials of the form
\begin{gather*}
\alpha(\lambda)= \alpha_1\lambda^{g-1} + \cdots + \alpha_g,
\\
\beta(\lambda) = \lambda^g + \beta_1\lambda^{g-1} + \cdots +
\beta_g,
\\
\gamma(\lambda) = \lambda^{g+1} + \gamma_1\lambda^g + \cdots +
\gamma_{g+1},
\end{gather*}
$h(\lambda)$ is a polynomial of the form $h(\lambda) =
\lambda^{2g+1} + \cdots$, and the spectral curve is a
hyperelliptic curve of genus $g$.

The spectral curve plays a central role in the algebro-geometric
theory of commuting pairs \mbox{\cite{DMN76,Krichever77}}. In that
case, the f\/lows of the KdV hierarchy can be translated to the
{\it isospectral} Lax equations
\begin{gather}
  \rd_{2n+1}V(\lambda) = [U_n(\lambda),V(\lambda)]
\label{Mumford}
\end{gather}
of the same matrix $V(\lambda)$ as we have used thus far (except
that $c_n$'s are genuine constant). In  particular, the polynomial
$h(\lambda)$ (hence the spectral curve itself) is invariant under
time evolutions:
\begin{gather*}
  \rd_{2n+1}h(\lambda) = 0.
\end{gather*}
This system \eqref{Mumford} is called `the Mumford system'
\cite{Mumford84}.

In contrast, the polynomial $h(\lambda)$ for the matrix Lax
equations \eqref{V-t-Laxeq} of the PI hierarchy is {\it not}
constant in time evolutions.  Straightforward calculations show
that the $t$-derivatives of $h(\lambda)$ take non-zero values as
\begin{gather*}
  \rd_{2n+1}h(\lambda) = \Tr U_n'(\lambda)V(\lambda).
\end{gather*}
For instance, the lowest equation for $n = 0$ ($t_1 = x$) reads
\begin{gather*}
  \rd_x h(\lambda) = \beta(\lambda).
\end{gather*}
The spectral curve thus deforms as $x$ and $t_{2n+1}$'s vary.
Obviously, the extra terms $U_n'(\lambda)$ are responsible for
this phenomena. This is a common feature of isomonodromic Lax
equations.

\subsection{Some technical lemmas}

As we shall show later on, the coef\/f\/icients of terms of higher
degrees in the polynomial $h(\lambda)$ have a rather special
structure. We present here a few technical lemmas that are used to
explain this fact.

The following lemmas are concerned with the KdV hierarchy rather
than the PI hierarchy. In the case of the KdV hierarchy, the
matrices $U_n(\lambda)$ are def\/ined for all nonnegative integers
$n = 0,1,2,\ldots$.

\begin{lemma}
There is a $2 \times 2$ matrix
\begin{gather*}
  \Phi(\lambda)
  = \left(\begin{matrix}
    1 + O(\lambda^{-1}) & O(\lambda^{-1}) \\
    w_1 + O(\lambda^{-1}) & 1 + O(\lambda^{-1})
    \end{matrix}\right)
\end{gather*}
of Laurent series of $\lambda$ that satisfies the equations
\begin{gather}
  \rd_{2n+1}\Phi(\lambda)
  = U_n(\lambda)\Phi(\lambda)
    - \Phi(\lambda)\lambda^n\Lambda,
  \qquad n = 0,1,2,\ldots,
\label{Phi-Sato-eq}
\end{gather}
where
\begin{gather*}
  \Lambda
  = \left(\begin{matrix}
    0 & 1 \\
    \lambda & 0
    \end{matrix}\right).
\end{gather*}
\end{lemma}

\proof Let $\psi(z)$ be the special solution, \eqref{normal-psi},
of the auxiliary linear equations.  In the case of the KdV
hierarchy, this is a function of the form $\psi(z) =
w(z)e^{\xi(z)}$, where
\begin{gather*}
  w(z) = 1 + \sum_{j=1}^\infty w_jz^{-j}, \qquad
  \xi(z) = \sum_{n=0}^\infty t_{2n+1}z^{2n+1}
  \qquad (t_1 = x).
\end{gather*}
The associated vector-valued function
\begin{gather*}
  \bspsi(z)
  = \left(\begin{matrix}
    \psi(z) \\
    \psi(z)_x
    \end{matrix}\right)
  = \left(\begin{matrix}
    w(z) \\
    zw(z) + w(z)_x
    \end{matrix}\right)
    e^{\xi(z)}
\end{gather*}
satisf\/ies the auxiliary linear equations
\begin{gather*}
  \rd_{2n+1}\bspsi(z) = U_n(\lambda)\bspsi(z)
   \qquad (\lambda = z^2)
\end{gather*}
with the same coef\/f\/icients $U_n(\lambda)$ as in
\eqref{bspsi-t-lineq} but now def\/ined by \eqref{U_n-def} for all
nonnegative integers $n = 0,1,2,\ldots$.

Moreover, since $\lambda$ remains invariant by substituting $z \to
-z$, $\bspsi(-z)$ is also a solution of these linear equations.
Thus we actually have a $2 \times 2$ matrix-valued solution
\begin{gather*}
  \bigl(\bspsi(z)\; \bspsi(-z)\bigl)
  = \left(\begin{matrix}
    w(z) & w(-z) \\
    zw(z) + w(z)_x & -zw(-z) + w(-z)_x
    \end{matrix}\right)
    \left(\begin{matrix}
    e^{\xi(z)} & 0 \\
    0 & e^{-\xi(z)}
    \end{matrix}\right).
\end{gather*}
We now consider
\begin{gather*}
\Psi(\lambda) = \bigl(\bspsi(z)\ \bspsi(-z)\bigr)
    \left(\begin{matrix}
    1 & 1 \\
    z & - z
    \end{matrix}\right)^{-1}
= \biggl(\frac{\bspsi(z) + \bspsi(-z)}{2} \
      \frac{\bspsi(z) - \bspsi(-z)}{2z} \biggr),
\end{gather*}
which is also a matrix-valued solution of the foregoing linear
equations. Note that the matrix elements are even functions of $z$
(hence functions of $\lambda$).   Moreover, $\Psi(\lambda)$ can be
factorized to the product of
\begin{gather*}
  \Phi(\lambda)
  = \left(\begin{matrix}
    w(z) & w(-z) \\
    zw(z) + w(z)_x & -zw(-z) + w(-z)_x
    \end{matrix}\right)
    \left(\begin{matrix}
    1 & 1 \\
    z & - z
    \end{matrix}\right)^{-1}
\end{gather*}
and
\begin{gather*}
  \left(\begin{matrix}
  1 & 1 \\
  z & - z
  \end{matrix}\right)
  \left(\begin{matrix}
  e^{\xi(z)} & 0 \\
  0 & e^{-\xi(z)}
  \end{matrix}\right)
  \left(\begin{matrix}
  1 & 1 \\
  z & - z
  \end{matrix}\right)^{-1}
  = \exp\Biggl(\sum_{n=0}^\infty
    t_{2n+1}\Lambda^{2n+1}\Biggr).
\end{gather*}
By plugging
\begin{gather*}
  \Psi(\lambda)
  = \Phi(\lambda)
    \exp\Biggl(\sum_{n=0}^\infty
      t_{2n+1}\Lambda^{2n+1}\Biggr)
\end{gather*}
into the auxiliary linear equations, $\Phi(\lambda)$ turns out to
satisfy the equations
\begin{gather*}
  \rd_{2n+1}\Phi(\lambda)
  = U_n(\lambda)\Phi(\lambda)
    - \Phi(\lambda)\Lambda^{2n+1},
\end{gather*}
which are nothing but \eqref{Phi-Sato-eq} because of the identity
$\Lambda^{2n+1} = \lambda^n\Lambda$. \qed

Now let us introduce the matrix
\begin{gather*}
  U(\lambda) = \Phi(\lambda)\Lambda\Phi(\lambda)^{-1}.
\end{gather*}
As a consequence of \eqref{Phi-Sato-eq}, it satisf\/ies the Lax
equations
\begin{gather}
 \rd_{2n+1}U(\lambda) = [U_n(\lambda),U(\lambda)],
  \qquad n = 0,1,2,\ldots.
\label{U-t-Laxeq}
\end{gather}
The following lemma shows that we can use $U(\lambda)$ as a kind
of generating function for $U_n(\lambda)$'s

\begin{lemma}
The matrix elements of
\begin{gather*}
  U(\lambda)
  = \left(\begin{matrix}
    \Alpha(\lambda) & \Beta(\lambda) \\
    \Gamma(\lambda) & - \Alpha(\lambda)
    \end{matrix}\right),
\end{gather*}
are Laurent series of $\lambda$ of the form
\begin{gather*}
  \Alpha(\lambda) = O(\lambda^{-1}), \qquad
  \Beta(\lambda) = 1 + O(\lambda^{-1}), \qquad
  \Gamma(\lambda) = \lambda + O(\lambda^{0}),
\end{gather*}
that satisfy the following algebraic conditions:
\begin{gather}
\Alpha_n(\lambda) = \bigl(\lambda^n\Alpha(\lambda)\bigr)_{\ge 0},
\qquad \Beta_n(\lambda) = \bigl(\lambda^n\Beta(\lambda)\bigr)_{\ge
0},\nonumber
\\
\Gamma_n(\lambda) = \bigl(\lambda^n\Gamma(\lambda)\bigr)_{\ge 0} -
R_{n+1}, \qquad n = 0,1,\ldots, \label{U_n-U}
\\
\Alpha(\lambda)^2  + \Beta(\lambda)\Gamma(\lambda) = \lambda,
\label{A^2+BC=l}
\end{gather}
where $(\ )_{\ge 0}$ stands for the polynomial part of a Laurent
series:
\begin{gather*}
  \left(\sum_{j\in\ZZ}a_j\lambda^j\right)_{\ge 0}
  = \sum_{j \ge 0} a_j\lambda^j.
\end{gather*}
\end{lemma}

\proof Rewrite \eqref{Phi-Sato-eq} as
\begin{gather*}
 U_n(\lambda)= \rd_{2n+1}\Phi(\lambda)\cdot\Phi(\lambda)^{-1}
    + \Phi(\lambda)\lambda^n\Lambda\Phi(\lambda)^{-1}
    = \rd_{2n+1}\Phi(\lambda)\cdot\Phi(\lambda)^{-1} + \lambda^n U(\lambda),
\end{gather*}
and compare the polynomial part of both hand sides.
$U_n(\lambda)$ is a matrix of polynomials, and the f\/irst term on
the right hand side is a matrix of the form
\begin{gather*}
  \rd_{2n+1}\Phi(\lambda)\cdot\Phi(\lambda)^{-1}
  = \left(\begin{matrix}
    O(\lambda^{-1}) & O(\lambda^{-1} \\
    \rd_{2n+1}w_1 + O(\lambda^{-1}) & O(\lambda^{-1})
    \end{matrix}\right).
\end{gather*}
Thus, also by recalling \eqref{w_1-t-eq}, \eqref{U_n-U} turns out
to hold. \eqref{A^2+BC=l} is an immediate consequence of the
def\/inition of $U(\lambda)$ and the identity $\det\Lambda =
\lambda$. \qed

\begin{remark}
Since $\Beta_n(\lambda)$ is equal to the auxiliary polynomial
$R_n(\lambda)$ def\/ined in \eqref{R_n(l)-def}, the second
equation of \eqref{U_n-U} implies that $\Beta(\lambda)$ is a
generating function of all $R_n$'s:
\begin{gather*}
  \Beta(\lambda)
  = 1 + R_1\lambda^{-1} + R_2\lambda^{-2} + \cdots.
\end{gather*}
The lowest ($n = 0$) part of \eqref{U-t-Laxeq} is not a genuine
evolution equation. In components, this equation reads
\begin{gather}
\rd_x\Alpha(\lambda)
  = \Gamma(\lambda) - (\lambda - u)\Beta(\lambda), \qquad
  \rd_x\Beta(\lambda) = - 2\Alpha(\lambda),\nonumber
  \\
\rd_x\Gamma(\lambda) = 2(\lambda - u)\Alpha(\lambda).
\label{U-x-Laxeq}
\end{gather}
The f\/irst two equations can be solved for $\Alpha(\lambda)$ and
$\Gamma(\lambda)$ as
\begin{gather*}
  \Alpha(\lambda) = - \frac{1}{2}\Beta(\lambda)_x, \qquad
  \Gamma(\lambda) = - \frac{1}{2}\Beta(\lambda)_{xx}
     + (\lambda - u)\Beta(\lambda).
\end{gather*}
The third equation thereby reduces to
\begin{gather}
  \frac{1}{2}\Beta(\lambda)_{xx}
  - 2(\lambda - u)\Beta(\lambda)_x
  + u_x\Beta(\lambda)
  = 0.
\label{R_n-genx}
\end{gather}
It is easy to see that this is a generating functional form of the
Lenard relations \eqref{Lenard}.
\end{remark}

\begin{remark}
If $\Alpha(\lambda)$ and $\Gamma(\lambda)$ are eliminated by
\eqref{U-x-Laxeq}, \eqref{A^2+BC=l} becomes another generating
functional formula
\begin{gather}
  \frac{1}{4}\bigl(\Beta(\lambda)_x\bigr)^2
  + \Beta(\lambda)\left(- \frac{1}{2}\Beta(\lambda)_{xx}
     + (\lambda - u)\Beta(\lambda) \right)
  = \lambda
\label{R_n-gen}
\end{gather}
of relations among $R_n$'s. \eqref{R_n-gen} may be thought of as a
once-integrated form of \eqref{R_n-genx}.  If expanded in powers
of $\lambda$, \eqref{R_n-gen} becomes a sequence of relations that
determine $R_n$ recursively {\it without integration procedure}.
This is an alternative and more practical way for calculating
$R_n$'s.
\end{remark}

\subsection[Detailed structure of $h(\lambda)$]{Detailed structure of $\boldsymbol{h(\lambda)}$}

We now turn to the issue of $h(\lambda)$. Let us recall
\eqref{V=U+...}. In components, it reads
\begin{gather*}
\alpha(\lambda)= \Alpha_g(\lambda) + c_1(t)\Alpha_{g-1}(\lambda)
     + \cdots + c_g(t)\Alpha_0(\lambda),
\\[1ex]
\beta(\lambda)= \Beta_g(\lambda) + c_1(t)\Beta_{g-1}(\lambda)
     + \cdots + c_g(t)\Beta_0(\lambda),
\\[1ex]
\gamma(\lambda)= \Gamma_g(\lambda) + c_1(t)\Gamma_{g-1}(\lambda)
     + \cdots + c_g(t)\Gamma_0(\lambda).
\end{gather*}
We can thereby express $h(\lambda)$ as
\begin{gather*}
  h(\lambda)
  = \sum_{m,n=0}^g c_m(t)c_n(t)
    (\Alpha_{g-m}(\lambda)\Alpha_{g-n}(\lambda)
    + \Beta_{g-m}(\lambda)\Gamma_{g-n}(\lambda)),
\end{gather*}
where it is understood that $c_0(t) = 1$. Moreover, since
\eqref{U_n-U} implies that
\begin{gather*}
\Alpha_n(\lambda)
  = \lambda^n\Alpha(\lambda) + O(\lambda^{-1}), \qquad
  \Beta_n(\lambda)
  = \lambda^n\Beta(\lambda) + O(\lambda^{-1}), \\[1ex]
\Gamma_n(\lambda)
  = \lambda^n\Gamma(\lambda) - R_{n+1} + O(\lambda^{-1}),
\end{gather*}
we can further rewrite $h(\lambda)$ as
\begin{gather*}
  h(\lambda)
  = \sum_{m,n=0}^g c_m(t)c_n(t)\lambda^{2g-m-n}
     (\Alpha(\lambda)^2 + \Beta(\lambda)\Gamma(\lambda)) \\
  \phantom{h(\lambda)=}
   - 2\sum_{m=0}^g c_m(t)R_{g+1-m}\lambda^g
   + O(\lambda^{g-1}).
\end{gather*}
We can now use \eqref{A^2+BC=l} and \eqref{higher-PI-eq} on the
right hand side. This leads to the following f\/inal result.

\begin{theorem}
Up to terms of $O(\lambda^{g-1})$, $h(\lambda)$ can be expressed
as
\begin{gather}
  h(\lambda)
  = \lambda^{2g+1} + 2c_1(t)\lambda^{2g}
    + (2c_2(t) + c_1(t)^2)\lambda^{2g-1}
    + (2c_3(t) + 2c_1(t)c_2(t))\lambda^{2g-2}+ \cdots\nonumber
\\ \phantom{h(\lambda)=}
   {}
    + \sum_{m=0}^g c_m(t)c_{g-m}(t)\lambda^{g+1}
    + \Biggl(\sum_{m=1}^{g-1}c_m(t)c_{g+1-m}(t)
      + x\Biggr)\lambda^g
    + O(\lambda^{g-1}).
\label{h-expansion}
\end{gather}
In particular, the coefficients of
$\lambda^{2g+1},\ldots,\lambda^g$ in $h(\lambda)$ do not contain
$u,u_x,\ldots$.
\end{theorem}

Let $I_0(\lambda)$ denote the part of $h(\lambda)$ consisting of
$\lambda^{2g+1}, \ldots,\lambda^g$, and $I_1,\ldots,I_g$ the
coef\/f\/icients of $\lambda^{g-1},\ldots,1$:
\begin{gather*}
  h(\lambda) = I_0(\lambda)
    + I_1\lambda^{g-1} + \cdots + I_g.
\end{gather*}
We have seen above that $I_0(\lambda)$ is a kinematical quantity
that is independent of the solution of the PI hierarchy in
question. In contrast, the remaining coef\/f\/icients
$I_1,\ldots,I_g$ are genuine dynamical quantities.

In the case of the Mumford system \eqref{Mumford}, these
coef\/f\/icients $I_1,\ldots,I_g$  are Hamiltonians of commuting
f\/lows.  More precisely, it is not these coef\/f\/icients but
their suitable linear combinations $H_1,\ldots,H_g$ that exactly
correspond to the f\/lows in $t_1,t_3,\ldots$. We shall encounter
the same problem in the case of the PI hierarchy.

\section{Hamiltonian structure of Lax equations}

We use the same Poisson structure as used for the Mumford system
\cite{Beauville90,NS01}. This Poisson structure is def\/ined on
the $3g+1$-dimensional moduli space of the matrix $V(\lambda)$
with coordinates $\alpha_1,\ldots,\alpha_g,\beta_1,\ldots,\beta_g,
\gamma_1,\ldots,\gamma_{g+1}$. It is customary to use the tensor
notation
\begin{gather*}
  \{V(\lambda) \stackrel{\otimes}{,} V(\mu)\}
  = \sum_{a,b,c,d=1,2}
    \{V_{ab}(\lambda),V_{cd}(\lambda)\}
    E_{ab} \otimes E_{cd},
\end{gather*}
where $E_{ab}$ denote the usual basis of $2 \times 2$ matrices.
The Poisson brackets of the matrix elements of $V(\lambda)$ can be
thereby written in a compact form as
\begin{gather*}
  \{V(\lambda) \stackrel{\otimes}{,} V(\mu)\}
= [V(\lambda) \otimes I \!+\! I \otimes V(\mu),
     r(\lambda \!-\! \mu))]
 \! + \![V(\lambda) \otimes I \!-\! I \otimes V(\mu),
     E_{21} \otimes E_{21}],
\end{gather*}
where $r(\lambda - \mu)$ is the standard rational $r$-matrix
\begin{gather*}
  r(\lambda - \mu) = \frac{P}{\lambda - \mu},\qquad
  P = \sum_{a,b=1,2} E_{ab} \otimes E_{ba}.
\end{gather*}
This is a version of the `generalized linear brackets'
\cite{EEKT94}.  More explicitly,
\begin{gather}
  \{\alpha(\lambda),\alpha(\mu)\} = 0,  \qquad
  \{\beta(\lambda),\beta(\mu)\} = 0,\nonumber
  \\[1ex]
  \{\alpha(\lambda),\beta(\mu)\}
  = \frac{\beta(\lambda) - \beta(\mu)}{\lambda - \beta},  \qquad
  \{\alpha(\lambda),\gamma(\mu)\}
  = - \frac{\gamma(\lambda) - \gamma(\mu)}{\lambda - \mu},\nonumber
  \\[1ex]
  \{\beta(\lambda),\gamma(\mu)\}
  = 2\frac{\alpha(\lambda) - \alpha(\mu)}{\lambda - \mu},  \qquad
  \{\gamma(\lambda),\gamma(\mu)\}
  = - 2\alpha(\lambda) + 2\alpha(\mu).
\label{Vcomp-PB}
\end{gather}

We can convert the Lax equations \eqref{V-t-Laxeq} of the PI
hierarchy to a Hamiltonian form with respect to this Poisson
structure. This procedure is fully parallel to the case of the
Mumford system.

A clue is the Poisson commutation relation
\begin{gather}
  \{V(\lambda),h(\mu)\}
   = \biggl[V(\lambda),\,
     \frac{V(\mu)}{\lambda - \mu}
     + \beta(\mu)E_{21}\biggr],
\label{V,h-PB}
\end{gather}
which can be derived from \eqref{Vcomp-PB} by straightforward
calculations. One can derive from this relation the Poisson
brackets $\{V(\lambda),I_{n+1}\}$ as follows.

\begin{lemma}
\begin{gather}
  \{V(\lambda),I_{n+1}\} = [V_n(\lambda),V(\lambda)],
\label{V,I_n-PB}
\end{gather}
where $V_n(\lambda)$ is a matrix of the form
\begin{gather*}
  V_n(\lambda)
  = \left(\begin{matrix}
    \alpha_n(\lambda) & \beta_n(\lambda) \\
    \gamma_n(\lambda) & - \alpha_n(\lambda)
    \end{matrix}\right)
\end{gather*}
with the matrix elements
\begin{gather*}
\alpha_n(\lambda) = \bigl(\lambda^{n-g}\alpha(\lambda)\bigr)_{\ge
0},  \qquad \beta_n(\lambda) =
\bigl(\lambda^{n-g}\beta(\lambda)\bigr)_{\ge 0},\nonumber
\\[1ex]
\gamma_n(\lambda) = \bigl(\lambda^{n-g}\gamma(\lambda)\bigr)_{\ge
0} - \beta_{n+1}, \qquad n = 0,1,\ldots,g-1.
%\label{V_n(l)-def}
\end{gather*}
\end{lemma}

\proof $I_{n+1}$ can be extracted from $h(\mu)$ by a contour
integral of the form
\begin{gather*}
  I_{n+1} = \oint\frac{d\mu}{2\pi i}\mu^{n-g}h(\mu),
\end{gather*}
where the contour is understood to be a circle around $\mu =
\infty$. The same contour integral applied to \eqref{V,h-PB}
yields the Poisson bracket in question:
\begin{gather*}
  \{V(\lambda),I_{n+1}\}= \oint\frac{d\mu}{2\pi i}\mu^{n-g}
     \{V(\lambda),h(\mu)\}= \biggl [V(\lambda), \oint\frac{d\mu}{2\pi i}
        \frac{\mu^{n-g}V(\mu)}{\lambda-\mu}  + \beta_{n+1}E_{21} \biggr].
\end{gather*}
Let us examine the matrix
\begin{gather*}
  V_n(\lambda)
  = - \oint\frac{d\mu}{2\pi i}
      \frac{\mu^{n-g}V(\mu)}{\lambda-\mu}
    - \beta_{n+1}E_{21}.
\end{gather*}
Since this type of contour integral gives, up to signature,  the
polynomial part of a Laurent series~$f(\lambda)$ as
\begin{gather*}
  \oint\frac{d\mu}{2\pi i}
  \frac{f(\mu)}{\lambda-\mu}
  = - \bigl(f(\lambda)\bigl)_{\ge 0},
\end{gather*}
$V_n(\lambda)$ can be expressed as
\begin{gather*}
  V_n(\lambda)
  = \bigl(\lambda^{n-g}V(\lambda)\bigr)_{\ge 0}
    - \beta_{n+1}E_{21}.
\end{gather*}
The statement of the lemma follows from this expression of
$V_n(\lambda)$. \qed

Actually, this result is not what we really want -- we have to
derive $[U_n(\lambda),V(\lambda)]$ rather than
$[V_n(\lambda),V(\lambda)]$. Here we need another clue, which is
the following linear relations among $U_n(\lambda)$'s and
$V_n(\lambda)$'s.

\begin{lemma}
\begin{gather}
  V_0(\lambda) = U_0(\lambda),\nonumber
  \\[1ex]
  V_n(\lambda) = U_n(\lambda) + c_1(t)U_{n-1}(\lambda)
    + \cdots + c_n(t)U_0(\lambda),
  \qquad n = 1,\ldots,g-1.
\label{V_n=U_n+...}
\end{gather}
\end{lemma}

\proof The linear relations \eqref{beta_n=R_n+...} among
$\beta_n$'s and $R_n$'s imply the linear relations
\begin{gather*}
\beta_n(\lambda)= R_n(\lambda) + c_1(t)R_{n-1}(\lambda)  + \cdots
+ c_n(t)R_0(\lambda)
\\ \phantom{\beta_n(\lambda)}{}
= \Beta_n(\lambda) + c_1(t)\Beta_{n-1}(\lambda)  + \cdots +
c_n(t)\Beta_0(\lambda)
\end{gather*}
of the auxiliary polynomials. On the other hand, $\alpha(\lambda)$
and $\gamma(\lambda)$ s are connected with $\beta(\lambda)$ as
\begin{gather*}
  \alpha(\lambda) = - \frac{1}{2}\beta(\lambda)_x,
  \qquad
  \gamma(\lambda)
  = - \frac{1}{2}\beta(\lambda)_{xx}
    + (\lambda - u)\beta(\lambda),
\end{gather*}
Expanding these relations in powers of $\lambda$ and picking out
the terms contained $\alpha_n(\lambda)$ and $\gamma_n(\lambda)$,
one can see that $\alpha_n(\lambda)$ and $\gamma_n(\lambda)$ are
linearly related to $\Alpha_n(\lambda)$'s and
$\Gamma_n(\lambda)$'s with the same coef\/f\/icients as{\samepage
\begin{gather*}
 \alpha_n(\lambda)= \Alpha_n(\lambda) + c_1(t)\Alpha_{n-1}(\lambda)
  + \cdots + c_n(t)\Alpha_0(\lambda),
\\[1ex]
\gamma_n(\lambda)= \Gamma_n(\lambda) + c_1(t)\Gamma_{n-1}(\lambda)
  + \cdots + c_n(t)\Gamma_0(\lambda).
\end{gather*}
These are exactly the linear relations presented in
\eqref{V_n=U_n+...} in a matrix form.} \qed

In view of this lemma, we def\/ine new Hamiltonians
$H_1,\ldots,H_g$ by the (triangular) linear equations
\begin{gather}
  I_1 = H_1,\qquad
  I_{n+1} = H_{n+1} + c_1(t)H_n
            + \cdots + c_n(t)H_1,
  \qquad n = 1,\ldots,g-1.
\label{H_n-def}
\end{gather}
Note that $H_{g+1}$ is not def\/ined (because $I_{g+1}$ does not
exist). The foregoing formula \eqref{V,I_n-PB} of the Poisson
brackets of $V(\lambda)$ and  $I_n$'s can be thereby converted to
the form
\begin{gather*}
  \{V(\lambda),H_{n+1}\}
  = [U_n(\lambda),V(\lambda)]
\end{gather*}
that we have sought for.  We thus eventually obtain the following
result.

\begin{theorem}
Except for the $t_{2g+1}$-flow, the matrix Lax equations
\eqref{V-t-Laxeq} of the PI hierarchy can be cast into the
Hamiltonian form
\begin{gather}
  \rd_{2n+1}V(\lambda)
  = \{V(\lambda),H_{n+1}\} + U_n'(\lambda),
  \qquad n = 0,1,\ldots,g-1,
\label{V-t-Hameq}
\end{gather}
with the Hamiltonians defined by \eqref{H_n-def}.
\end{theorem}

\begin{remark}
As regards the excluded $t_{2g+1}$-f\/low, the polynomial
$h(\lambda)$ obviously contains no candidate of Hamiltonian. If we
naively extrapolate \eqref{H_n-def} to $n = g$, we end up with the
linear relation
\begin{gather*}
  0= I_{g+1}
   = H_{g+1} + c_1(t)H_g + \cdots + c_g(t)H_1
\end{gather*}
of the Hamiltonians.  In a sense, this is a correct statement,
which says that $H_{g+1}$ is not an independent Hamiltonian.
\end{remark}

\begin{remark}
$I_0(\lambda)$ is a central element (i.e., a Casimir function) of
the Poisson algebra. To see this, note that the right hand side of
\eqref{V,h-PB} is of order $O(\mu^{g-1})$ as $\mu \to \infty$.
This implies that the terms of degree greater than $g$ in $h(\mu)$
have no contribution to $\{V(\lambda),h(\mu)\}$, in other words,
\begin{gather}
  \{V(\lambda),I_0(\mu)\} = 0.
\label{V,I_0-PB}
\end{gather}
This means that $I_0(\mu)$ is a Casimir function. This fact is in
accord with the observation in the last section that
$I_0(\lambda)$ does not contain genuine dynamical variables.
\end{remark}

\begin{remark}
$I_n$'s are Poisson-commuting, i.e., $\{I_j,I_k\} = 0$.  This is a
consequence of another basic Poisson relation
\begin{gather*}
  \{h(\lambda),h(\mu)\} = 0,
\end{gather*}
which, too, can be derived from \eqref{Vcomp-PB} by
straightforward calculations (or by a standard $r$-matrix
technique).
\end{remark}

\section{Spectral Darboux coordinates}

The construction of `Spectral Darboux coordinates' is also
parallel to the case of the Mumford system. These coordinates
consist of the roots $\lambda_1,\ldots,\lambda_g$ of
$\beta(\lambda)$ and the values $\mu_1,\ldots,\mu_g$ of
$\alpha(\lambda)$ at these roots of $\beta(\lambda)$:
\begin{gather*}
  \beta(\lambda) = \prod_{j=1}^g (\lambda - \lambda_j),
  \qquad
  \mu_j = \alpha(\lambda_j), \qquad j = 1,\ldots,g.
\end{gather*}
To avoid delicate problems, the following consideration is limited
to a domain of the phase space where $\lambda_j$'s are distinct.

$\lambda_j$ and $\mu_j$ satisfy the equation of the spectral
curve:
\begin{gather*}
  \mu_j^2 = h(\lambda_j).
%\label{m_j^2=h(l_j)}
\end{gather*}
We thus have a $g$-tuple $(\lambda_j,\mu_j)_{j=1}^g$ of points of
the spectral curve (in other words, an ef\/fective divisor of
degree $g$) that represents a point of the Jacobi variety of the
spectral curve.  In the case of the Mumford system, the commuting
f\/lows are thereby mapped to linear f\/lows on the Jacobi variety
\cite{DMN76,Krichever77, Mumford84}.  The case of the PI hierarchy
is more complicated because the spectral curve itself is
dynamical.  If one wishes to pursue this approach, one has to
consider the coupled dynamics of both the divisor and the
underlying spectral curve; unlike the case of isospectral
problems, this does not reduce the complexity of dynamics of the
original nonlinear problem.  Actually, this is not what we seek
for.  We simply borrow the idea of spectral Darboux coordinates to
describe the Hamiltonian structure of the system in question.

As it follows from \eqref{Vcomp-PB} by a standard procedure
\cite{EEKT94,AHH93,Sklyanin95}, these new variables satisfy the
canonical Poisson relations
\begin{gather*}
  \{\lambda_j,\lambda_k\} = 0, \qquad
  \{\mu_j,\mu_k\} = 0, \qquad
  \{\lambda_j,\mu_k\} = \delta_{jk}.
\end{gather*}
On the other hand, they Poisson-commute with $I_0(\lambda)$,
\begin{gather*}
  \{\lambda_j,I_0(\lambda)\} = 0, \qquad
  \{\mu_j,I_0(\lambda)\} = 0,
\end{gather*}
because $I_0(\lambda)$ is a Casimir function as \eqref{V,I_0-PB}
shows.  Thus $\lambda_j$'s and $\mu_j$'s may be literally called
`Darboux coordinates'.

These Darboux coordinates $\lambda_j,\mu_j$ and the
coef\/f\/icients of $\lambda^{2g},\ldots, \lambda^g$ in
$I_0(\lambda) = \lambda^{2g+1} + \cdots$ give an alternative
(local) coordinate system of the $3g+1$-dimensional Poisson
structure on the space of $L$-matrices with the original
$g+g+(g+1)$ coordinates $\gamma_j$, $\beta_j$, $\gamma_j$.  To
reconstruct the $L$-matrix $V(\lambda)$ from these new
coordinates, we use the familiar Lagrange interpolation formula
\begin{gather}
  f(\lambda)
  = \sum_{j=1}^g
    \frac{f(\lambda_j)}{\beta'(\lambda_j)}
    \frac{\beta(\lambda)}{\lambda - \lambda_j}
\label{Lagrange1}
\end{gather}
that holds for any polynomial $f(\lambda) = f_1\lambda^{g-1} +
\cdots + f_g$ of degree less than $g$.  Since
\begin{gather*}
  \frac{\beta(\lambda)}{\lambda - \lambda_j}
  = - \frac{\rd\beta(\lambda)}{\rd\lambda_j}
  = - \sum_{n=1}^g
      \frac{\rd\beta_n}{\rd\lambda_j}\lambda^{g-n},
\end{gather*}
this formula implies the formula
\begin{gather}
  f_n = - \sum_{j=1}^g
          \frac{f(\lambda_j)}{\beta'(\lambda_j)}
          \frac{\rd\beta_n}{\rd\lambda_j}
\label{Lagrange2}
\end{gather}
for the coef\/f\/icients of $f(\lambda)$ as well. Note that
$\beta_n$'s are understood here to be functions of $\lambda_j$'s
(in fact, they are elementary symmetric functions). We apply this
formula \eqref{Lagrange2} to $\alpha(\lambda)$ and obtain the
explicit formula
\begin{gather*}
  \alpha_n
  = - \sum_{j=1}^g
      \frac{\mu_j}{\beta'(\lambda_j)}
      \frac{\rd\beta_n}{\rd\lambda_j}
%\label{alpha_n(l,m)}
\end{gather*}
that recovers $\alpha_n$'s from $\lambda_j$'s and $\mu_j$'s. In a
similar way, we apply \eqref{Lagrange2} to the case where
\begin{gather*}
  f(\lambda)
  = \sum_{n=1}^n I_n\lambda^{g-n}
  = h(\lambda) - I_0(\lambda),
\end{gather*}
and f\/ind the expression
\begin{gather}
  I_n = - \sum_{j=1}^g
      \frac{\mu_j^2 - I_0(\lambda_j)}{\beta'(\lambda_j)}
      \frac{\rd\beta_n}{\rd\lambda_j}
\label{I_n(l,m)}
\end{gather}
of $I_n$'s in terms of $\lambda_j$'s, $\mu_j$'s and
$I_0(\lambda)$.  For instance,
\begin{gather*}
  I_1 = \sum_{j=1}^g
        \frac{\mu_j^2 - I_0(\lambda_j)}{\beta'(\lambda_j)}.
\end{gather*}
In these formulas, $I_0(\lambda)$ is understood to be the
polynomial
\begin{gather*}
  I_0(\lambda)= \lambda^{2g+1} + 2c_1(t)\lambda^{2g}
    + (2c_2(t) + c_1(t)^2)\lambda^{2g-1}
    + (2c_3(t) + 2c_1(t)c_2(t))\lambda^{2g-2}+ \cdots \nonumber
 \\ \phantom{I_0(\lambda)=}
    {}
    + \sum_{m=0}^g c_m(t)c_{g-m}(t)\lambda^{g+1}
    + \Biggl(\sum_{m=1}^{g-1}c_m(t)c_{g+1-m}(t)
      + x\Biggr)\lambda^g.
%\label{I_0-redef}
\end{gather*}
Once $\alpha(\lambda)$ and $h(\lambda)$ are thus reconstructed, we
can recover $\gamma(\lambda)$ as
\begin{gather*}
  \gamma(\lambda)
  = \frac{h(\lambda) - \alpha(\lambda)^2}{\beta(\lambda)}.
\end{gather*}

It is convenient to rewrite the foregoing formula \eqref{I_n(l,m)}
slightly. Recall the auxiliary polynomials
\begin{gather*}
  \beta_n(\lambda)
  = \lambda^n + \beta_1\lambda^{n-1}
    + \cdots + \beta_n.
\end{gather*}

\begin{lemma}
\begin{gather}
  \frac{\rd\beta_n}{\rd\lambda_j}
  = - \beta_{n-1}(\lambda_j),
  \qquad n = 1,\ldots,g.
\label{dbeta_n}
\end{gather}
\end{lemma}

\proof Start from the identity
\begin{gather*}
  \frac{\rd\beta(\lambda)}{\rd\lambda_j}
  = - \frac{\beta(\lambda)}{\lambda - \lambda_j}
  = - \frac{\beta(\lambda) - \beta(\lambda_j)}
           {\lambda - \lambda_j}
\end{gather*}
and do substitution
\begin{gather*}
  - \beta_n\frac{\lambda^n - \lambda_j^n}{\lambda - \lambda_j}
  = - \beta_n(\lambda^{n-1} + \lambda_j\lambda^{n-2}
        + \cdots + \lambda_j^{n-2}\lambda + \lambda_j^{n-1})
\end{gather*}
for each term on the right hand side. This leads to  the identity
\begin{gather*}
  \frac{\rd\beta(\lambda)}{\rd\lambda_j}
 = - \lambda^{g-1}
     - (\lambda_j + \beta_1)\lambda^{g-2}
     - \cdots
     - (\lambda_j^{g-1} + \beta_1\lambda_j^{g-2}
        + \cdots + \beta_{g-1})
 \\ \qquad
 = - \lambda^{g-1}
     - \beta_1(\lambda_j)\lambda^{g-2}
     - \cdots - \beta_{g-1}(\lambda_j),
\end{gather*}
which implies \eqref{dbeta_n}. \qed

By these identities, we can rewrite \eqref{I_n(l,m)} as
\begin{gather*}
  I_{n+1}
  = \sum_{j=1}^g
    \frac{\mu_j^2 - I_0(\lambda_j)}{\beta'(\lambda_j)}
    \beta_n(\lambda_j),
  \qquad n = 0,1,\ldots,g-1.
%\label{I_n(l,m)-alt}
\end{gather*}
(For notational convenience, $n$ is shifted by one). We can
derive, from these formulas, a similar expression of the
Hamiltonians $H_{n+1}$ introduced in the last section. Recall that
$\beta_n$'s are connected with $R_n$'s by the linear relation
\eqref{beta_n=R_n+...}.  It is easy to see that the auxiliary
polynomials $\beta_n(\lambda)$, too, are linearly related to the
auxiliary polynomials $R_n(\lambda)$ as
\begin{gather*}
  \beta_n(\lambda)
  = R_n(\lambda) + c_1(t)R_{n-1}(\lambda)
    + \cdots + c_n(t)R_0(\lambda).
%\label{beta_n(l)=R_n(l)+...}
\end{gather*}
Comparing this linear relation with the linear relation
\eqref{H_n-def} among $I_n$'s and $H_n$'s, we f\/ind that the
Hamiltonians $H_{n+1}$ can be expressed as
\begin{gather*}
  H_{n+1}
  = \sum_{j=1}^g
    \frac{\mu_j^2 - I_0(\lambda_j)}{\beta'(\lambda_j)}
    R_n(\lambda_j),
  \qquad n = 0,1,\ldots,g-1.
%\label{H_n(l,m)}
\end{gather*}
Note that $R_n$'s in this formula have to be redef\/ined as
functions of $\lambda_j$ that satisfy the linear relations
\eqref{beta_n=R_n+...}:
\begin{gather*}
  R_1 = \beta_1 - c_1(t),\qquad
  R_2 = \beta_2 - c_1(t)\beta_1 + c_1(t)^2 - c_2(t),
  \quad \ldots.
\end{gather*}
In particular, these Hamiltonians are time-dependent, the
time-dependence stemming from both $I_0(\lambda)$ and
$R_n(\lambda)$.

This is, however, not the end of the story. As it turns out below,
these Hamiltonians (except $H_1 = I_1$) do {\it not} give correct
equations of motion in the Darboux coordinates $\lambda_j$,
$\mu_j$. Correct Hamiltonians are obtained by adding correction
terms to $H_n$'s.

\section{Hamiltonians in Darboux coordinates}

\subsection{Equations of motion in Darboux coordinates}

Let us derive equations of motion for $\lambda_j$'s and $\mu_j$'s
from the Lax equations \eqref{V-t-Laxeq}. In components, the Lax
equations take the following form:
\begin{gather}
  \rd_{2n+1}\alpha(\lambda)
  = \Beta_n(\lambda)\gamma(\lambda)
    - \beta(\lambda)\Gamma_n(\lambda)
    + \Alpha_n'(\lambda),\nonumber
    \\[.5ex]
  \rd_{2n+1}\beta(\lambda)
  = 2\Alpha_n(\lambda)\beta(\lambda)
    - 2\Beta_n(\lambda)\alpha(\lambda)
    + \Beta_n'(\lambda),\nonumber
    \\[.5ex]
  \rd_{2n+1}\gamma(\lambda)
  = 2\Gamma_n(\lambda)\alpha(\lambda)
    - 2\Alpha_n(\lambda)\gamma(\lambda)
    + \Gamma_n'(\lambda).
\label{V-t-Laxeq2}
\end{gather}

To derive equations of motion for $\lambda_j$'s, we
dif\/ferentiate the identity $\beta(\lambda_j) = 0$ by $t_{2n+1}$.
By the chain rule, this yields the equation
\begin{gather*}
  \rd_{2n+1}\beta(\lambda)|_{\lambda=\lambda_j}
  + \beta'(\lambda_j)\rd_{2n+1}\lambda_j
  = 0.
\end{gather*}
By the second equation of \eqref{V-t-Laxeq2}, the f\/irst term on
the right hand side can be expressed as
\begin{gather*}
  \rd_{2n+1}\beta(\lambda)|_{\lambda=\lambda_j}
  = - 2\Beta_n(\lambda)\alpha(\lambda_j)
    + \Beta_n'(\lambda_j)
  = - 2\mu_j\Beta_n(\lambda)
    + \Beta_n'(\lambda_j).
\end{gather*}
Thus the following equations are obtained for $\lambda_j$'s:
\begin{gather}
  \rd_{2n+1}\lambda_j
  = \frac{2\mu_j\Beta_n(\lambda_j)}{\beta'(\lambda_j)}
    - \frac{\Beta_n'(\lambda_j)}{\beta'(\lambda_j)}.
\label{lam-t-eq}
\end{gather}

To derive equations of motion for $\mu_j$'s, we dif\/ferentiate
$\mu_j = \alpha(\lambda_j)$ by $t_{2n+1}$. The outcome is the
equation
\begin{gather*}
  \rd_{2n+1}\mu_j
  = \rd_{2n+1}\alpha(\lambda)|_{\lambda=\lambda_j}
    + \alpha'(\lambda_j)\rd_{2n+1}\lambda_j.
\end{gather*}
By the f\/irst equation of \eqref{V-t-Laxeq2}, the f\/irst term on
the right hand side can be expressed as
\begin{gather*}
  \rd_{2n+1}\alpha(\lambda)|_{\lambda=\lambda_j}
  = \Beta_n(\lambda_j)\gamma(\lambda_j)
    + \Alpha_n'(\lambda_j).
\end{gather*}
The derivative $\rd_{2n+1}\lambda_j$ in the second term can be
eliminated by \eqref{lam-t-eq}. We can thus rewrite the foregoing
equation as
\begin{gather*}
  \rd_{2n+1}\mu_j
  = \frac{2\mu_j\alpha'(\lambda_j)
      + \beta'(\lambda_j)\gamma(\lambda_j)}
         {\beta'(\lambda_j)}\Beta_n(\lambda_j)
    - \frac{\alpha'(\lambda_j)\Beta_n'(\lambda_j)}
       {\beta'(\lambda_j)}
    + \Alpha_n'(\lambda_j).
\end{gather*}
Note here that the numerator of the f\/irst term on the right hand
side is just the value of $h'(\lambda)$ at $\lambda =\lambda_j$:
\begin{gather*}
  h'(\lambda_j)
  = 2\alpha(\lambda_j)\alpha'(\lambda_j)
    + \beta'(\lambda_j)\gamma(\lambda_j)
    + \beta(\lambda_j)\gamma'(\lambda_j)
  = 2\mu_j\alpha'(\lambda_j)
    + \beta'(\lambda_j)\gamma(\lambda_j).
\end{gather*}
Thus the following equations of motion are obtained for $\mu_j$'s:
\begin{gather}
  \rd_{2n+1}\mu_j
  = \frac{h'(\lambda_j)\Beta_n(\lambda_j)}{\beta'(\lambda_j)}
    - \frac{\alpha'(\lambda_j)\Beta_n'(\lambda_j)}
       {\beta'(\lambda_j)}
    + \Alpha_n'(\lambda_j).
\label{mu-t-eq}
\end{gather}

\subsection{Why Hamiltonians need corrections}

We now start from the Hamiltonian form \eqref{V-t-Hameq} of the
Lax equations and repeat similar calculations. In the case of
isospectral Lax equations, such as the Mumford system, this
procedure should lead to a Hamiltonian form of equations of motion
for the spectral Darboux coordinates.  In components,
\eqref{V-t-Hameq} consist of the following three sets of
equations:
\begin{gather}
  \rd_{2n+1}\alpha(\lambda)
  = \{\alpha(\lambda),H_{n+1}\} + \Alpha_n'(\lambda),\nonumber
  \\[1ex]
  \rd_{2n+1}\beta(\lambda)
  = \{\beta(\lambda),H_{n+1}\} + \Beta_n'(\lambda), \nonumber
  \\[1ex]
  \rd_{2n+1}\gamma(\lambda)
  = \{\gamma(\lambda),H_{n+1}\} + \Gamma_n'(\lambda).
\label{V-t-Hameq2}
\end{gather}

We again start from the identity
\begin{gather*}
  0 = \rd_{2n+1}\beta(\lambda_j)
    = \rd_{2n+1}\beta(\lambda)|_{\lambda=\lambda_j}
    + \beta'(\lambda_j)\rd_{2n+1}\lambda_j
\end{gather*}
and consider the second equation of \eqref{V-t-Hameq2}, which
implies that
\begin{gather*}
  \rd_{2n+1}\beta(\lambda)|_{\lambda=\lambda_j}
  = \{\beta(\lambda),H_{n+1}\}|_{\lambda=\lambda_j}
    + \Beta_n'(\lambda_j).
\end{gather*}
To calculate the f\/irst term on the right hand side, we use the
identity
\begin{gather*}
  0 = \{\beta(\lambda_j),H_{n+1}\}
    = \{\beta(\lambda),H_{n+1}\}|_{\lambda=\lambda_j}
      + \beta'(\lambda_j)\{\lambda_j,H_{n+1}\}.
\end{gather*}
Thus the following equations of motion are obtained for
$\lambda_j$'s:
\begin{gather}
  \rd_{2n+1}\lambda_j
  = \{\lambda_j,H_{n+1}\}
    - \frac{\Beta_n'(\lambda_j)}{\beta'(\lambda_j)}
\label{lam-t-eq2}
\end{gather}

In much the same way, we can derive the following equations of
motion for $\mu_j$'s:
\begin{gather}
  \rd_{2n+1}\mu_j
  = \{\mu_j,H_{n+1}\}
    - \frac{\alpha'(\lambda_j)\Beta_n'(\lambda_j)}
       {\beta'(\lambda_j)}
    + \Alpha_n'(\lambda_j)
\label{mu-t-eq2}
\end{gather}

These results clearly show that $H_{n+1}$ is not a correct
Hamiltonian. If we could f\/ind a correct Hamiltonian, say
$K_{n+1}$, the equations of motion would take the canonical form
\begin{gather*}
  \rd_{2n+1}\lambda_j = \{\lambda_j,K_{n+1}\},
  \qquad
  \rd_{2n+1}\mu_j = \{\mu_j,K_{n+1}\}.
\end{gather*}
\eqref{lam-t-eq2} and \eqref{mu-t-eq2} fail to take this form
because of extra terms on the right hand side. These terms stem
from the extra term $U_n'(\lambda)$ in the Lax equations
\eqref{V-t-Laxeq}.  When we say \eqref{V-t-Laxeq} is
`Hamiltonian', we ignore the presence of this term. It, however,
cannot be ignored if we attempt to formulate the Hamiltonian
structure in the language of the spectral Darboux coordinates
$\lambda_j$ and $\mu_j$.

The case of $n = 0$ is exceptional. Since $\Beta_0(\lambda) = 1$,
the extra terms on the right hand side of~\eqref{lam-t-eq} and
\eqref{mu-t-eq} disappear.  Therefore
\begin{gather*}
  H_1 = I_1
  = \sum_{j=1}^g
    \frac{\mu_j^2 - I_0(\lambda_j)}{\beta'(\lambda_j)}
%\label{H_1}
\end{gather*}
is a correct Hamiltonian for the equations of motion for the
$t_1$-f\/low (namely, the higher order PI equation itself). Note,
however, that the extra term $U_0'(\lambda)$ in the Lax equation
still persist.

\begin{remark}
Another consequence of the foregoing calculations is that the
Poisson brackets of~$H_{n+1}$ and the Darboux coordinates (in
other words, the components of the Hamiltonian vector f\/ield of
$H_{n+1}$) are given by
\begin{gather*}
  \{\lambda_j,H_{n+1}\}
  = \frac{2\mu_j\Beta_n(\lambda_j)}{\beta'(\lambda_j)},
  \qquad
  \{\mu_j,H_{n+1}\}
  = \frac{h'(\lambda_j)\Beta_n(\lambda_j)}{\beta'(\lambda_j)}.
\end{gather*}
Consequently, the Poisson brackets of $I_{n+1}$ and the Darboux
coordinates turn out to be given~by
\begin{gather*}
  \{\lambda_j,I_{n+1}\}
  = \frac{2\mu_j\beta_n(\lambda_j)}{\beta'(\lambda_j)},
  \qquad
  \{\mu_j,I_{n+1}\}
  = \frac{h'(\lambda_j)\beta_n(\lambda_j)}{\beta'(\lambda_j)}.
\end{gather*}
One can derive these results directly from the Poisson brackets
\eqref{Vcomp-PB} of the matrix elements of~$V(\lambda)$ as well.
\end{remark}

\subsection{Corrected Hamiltonians}

We now seek for correct Hamiltonians in such a form as
\begin{gather*}
  K_{n+1} = H_{n+1} + \Delta H_{n+1}.
\end{gather*}
Correction terms $\Delta H_{n+1}$ have to be chosen to satisfy the
conditions
\begin{gather*}
  \{\lambda_j,\Delta H_{n+1}\}
  = - \frac{\Beta_n'(\lambda_j)}{\beta'(\lambda_j)},
  \qquad
  \{\mu_j,\Delta H_{n+1}\}
  = - \frac{\alpha'(\lambda_j)\Beta_n'(\lambda_j)}
        {\beta'(\lambda_j)}
    + \Alpha_n'(\lambda_j).
\end{gather*}

We can convert this problem to that of $I_{n+1}$'s, namely, the
problem to identify the correction term to $I_{n+1}$'s:
\begin{gather}
  \Delta I_{n+1}
  = \Delta H_{n+1} + c_1(t)\Delta H_n
    + \cdots + c_{n-1}(t) \Delta H_1.
\label{DI_n=DH_n+...}
\end{gather}
The conditions for to $\Delta I_{n+1}$ read
\begin{gather*}
  \{\lambda_j,\Delta I_{n+1}\}
  = - \frac{\beta_n'(\lambda_j)}{\beta'(\lambda_j)},
  \qquad
  \{\mu_j,\Delta I_{n+1}\}
  = - \frac{\alpha'(\lambda_j)\beta_n'(\lambda_j)}
        {\beta'(\lambda_j)}
    + \alpha_n'(\lambda_j)
\end{gather*}
or, equivalently, {\samepage\begin{gather}
  \frac{\rd\Delta I_{n+1}}{\rd\mu_j}
  = - \frac{\beta_n'(\lambda_j)}{\beta'(\lambda_j)},
  \qquad
  \frac{\rd\Delta I_{n+1}}{\rd\lambda_j}
  = \frac{\alpha'(\lambda_j)\beta_n'(\lambda_j)}
      {\beta'(\lambda_j)}
   - \alpha_n'(\lambda_j).
\label{DI_n-condition}
\end{gather}
This slightly simplif\/ies the nature of the problem.

}

The goal of the subsequent consideration is to prove that a
correct answer to this question is given by
\begin{gather}
  \Delta I_{n+1}
  = - \sum_{k=1}^g
      \frac{\mu_k\beta_n'(\lambda_k)}{\beta'(\lambda_k)}.
\label{DI_n-answer}
\end{gather}
Obviously, the f\/irst half of \eqref{DI_n-condition} is
satisf\/ied; what remains is to check the second half.

We can use the following lemma to reduce the problem to each term
of the sum in \eqref{DI_n-answer}.

\begin{lemma}
\begin{gather*}
  \alpha'(\lambda_j)
  = - \sum_{k=1}^g \frac{\mu_k}{\beta'(\lambda_k)}
       \frac{\rd\beta'(\lambda)}{\rd\lambda_k}
       \bigg|_{\lambda=\lambda_j},
\qquad
  \alpha'_n(\lambda_j)
 = - \sum_{k=1}^g \frac{\mu_k}{\beta_n'(\lambda_k)}
      \frac{\rd\beta'_n(\lambda)}{\rd\lambda_k}
      \bigg|_{\lambda=\lambda_j}.
\end{gather*}
\end{lemma}

\proof By the Lagrange interpolation formula \eqref{Lagrange1},
$\alpha(\lambda)$ can be expressed as
\begin{gather*}
  \alpha(\lambda)
  = - \sum_{k=1}^g \frac{\mu_k}{\beta'(\lambda_k)}
        \frac{\rd\beta(\lambda)}{\rd\lambda_k}.
\end{gather*}
Applying the projection operator $(\lambda^{n-g}\;\cdot\;)$ to
both hand sides yields another identity
\begin{gather*}
  \alpha_n(\lambda)
  = - \sum_{k=1}^g \frac{\mu_k}{\beta_n'(\lambda_k)}
        \frac{\rd\beta_n(\lambda)}{\rd\lambda_k}.
\end{gather*}
The statement of the lemma follows by dif\/ferentiating both hand
sides of these identities and by setting $\lambda = \lambda_j$.
\qed

Because of this lemma, checking the second half of
\eqref{DI_n-condition} can be reduced to proving the following
identity:
\begin{gather}
  \frac{\rd}{\rd\lambda_j}
  \frac{\beta_n'(\lambda_k)}{\beta'(\lambda_k)}
  = \frac{1}{\beta'(\lambda_k)}
    \left.\left(
        \frac{\beta_n'(\lambda_j)}{\beta'(\lambda_j)}
        \frac{\rd\beta'(\lambda)}{\rd\lambda_k}
      - \frac{\rd\beta_n'(\lambda)}{\rd\lambda_k}
    \right)\right|_{\lambda=\lambda_j}
\label{DI_n-goal}
\end{gather}
Note that this is a genuinely algebraic problem (related to
elementary symmetric functions). We prepare some technical lemmas.

\begin{lemma}
\begin{gather}
  \frac{\rd^2\beta_n}{\rd\lambda_j\rd\lambda_k}
  = - \frac{1}{\lambda_j - \lambda_k}
      \biggl(\frac{\rd\beta_n}{\rd\lambda_j}
          - \frac{\rd\beta_n}{\rd\lambda_k}\biggr)
  \qquad (j \not= k), \qquad
  \frac{\rd^2\beta_n}{\rd\lambda_k^2}
  = 0 .
\label{d^2beta_n}
\end{gather}
\end{lemma}

\proof Dif\/ferentiate
\begin{gather*}
  \frac{\rd\beta(\lambda)}{\rd\lambda_k}
  = - \prod_{l\not=k}(\lambda - \lambda_l)
\end{gather*}
once again by $\lambda_j$. If $j \not= k$, this yields the
identity
\begin{gather*}
  \frac{\rd^2\beta(\lambda)}{\rd\lambda_j\rd\lambda_k}
 = \prod_{l\not=j,k}(\lambda - \lambda_l)
   = \frac{\beta(\lambda)}
     {(\lambda-\lambda_j)(\lambda-\lambda_k)}
= - \frac{1}{\lambda_j - \lambda_k}
       \biggl(\frac{\rd\beta(\lambda)}{\rd\lambda_j}
           - \frac{\rd\beta(\lambda)}{\rd\lambda_k}\biggr),
\end{gather*}
proving the f\/irst part of \eqref{d^2beta_n}. Similarly, if $j =
k$, the outcome is the identity $
  {\rd^2\beta(\lambda)}/{\rd\lambda_k^2} = 0,
$ which implies the rest of \eqref{d^2beta_n}. \qed

\begin{lemma}
\begin{gather}
  \beta_n'(\lambda_k)
  = - \frac{\rd\beta_n}{\rd\lambda_k}
    - \frac{\rd\beta_{n-1}}{\rd\lambda_k}\lambda_k
    - \cdots
    - \frac{\rd\beta_1}{\rd\lambda_k}\lambda_k^{n-1}.
\label{beta_n'(l_k)}
\end{gather}
\end{lemma}

\proof $\beta_n'(\lambda)$ can be expressed as
\begin{gather*}
 \beta_n'(\lambda)= n\lambda^{n-1} + (n-1)\beta_1\lambda^{n-2}
     + \cdots + \beta_{n-1}
\\ \phantom{\beta_n'(\lambda)}{}
= (\lambda^{n-1} + \beta_1\lambda^{n-2}
       + \cdots + \beta_{n-1})
     + \cdots + (\lambda + \beta_1)\lambda^{n-2}
     + \lambda^{n-1}
 \\ \phantom{\beta_n'(\lambda)}{}
= \beta_{n-1}(\lambda) + \beta_{n-2}(\lambda)\lambda
     + \cdots + \beta_0(\lambda).
\end{gather*}
Upon substituting $\lambda = \lambda_k$ and recalling
\eqref{dbeta_n}, \eqref{beta_n'(l_k)} follows. \qed

\begin{lemma}
\begin{gather}
  \frac{\rd\beta_n'(\lambda_k)}{\rd\lambda_j}
 = \frac{1}{\lambda_j - \lambda_k}
    \biggl(\frac{\rd\beta_n(\lambda_k)}{\rd\lambda_j}
      + \beta_n'(\lambda_k)\biggr)
  \qquad (j \not= k), \nonumber
  \\
  \frac{\rd\beta_n'(\lambda_k)}{\rd\lambda_k}
 = \frac{1}{2}\beta_n''(\lambda_k).
\label{dbeta_n'(l_k)}
\end{gather}
\end{lemma}

\proof If $j \not= k$, \eqref{beta_n'(l_k)} and \eqref{d^2beta_n}
imply that
\begin{gather*}
  \frac{\rd\beta_n'(\lambda_k)}{\rd\lambda_j}
  = - \frac{\rd^2\beta_n}{\rd\lambda_j\rd\lambda_k}
     - \frac{\rd^2\beta_{n-1}}{\rd\lambda_j\rd\lambda_k}
       \lambda_k
     - \cdots
     - \frac{\rd^2\beta_1}{\rd\lambda_j\rd\lambda_k}
       \lambda_k^{n-1}
 \\ \phantom{\frac{\rd\beta_n'(\lambda_k)}{\rd\lambda_j}}{}
   =  \frac{1}{\lambda_j-\lambda_k}
     \biggl(\frac{\rd\beta_n}{\rd\lambda_j}
         - \frac{\rd\beta_n}{\rd\lambda_k}\biggr)+ \cdots + \biggl(\frac{\rd\beta_1}{\rd\lambda_j}
         - \frac{\rd\beta_1}{\rd\lambda_k}\biggr)
       \lambda_k^{n-1}\\
 \phantom{\frac{\rd\beta_n'(\lambda_k)}{\rd\lambda_j}}{}
 = \frac{1}{\lambda_j-\lambda_k}
     \biggl(\frac{\rd\beta_n}{\rd\lambda_j}
         + \frac{\rd\beta_{n-1}}{\rd\lambda_j}\lambda_k
         + \cdots
         + \frac{\rd\beta_1}{\rd\lambda_j}\lambda_k^{n-1}\biggr)
\\ \phantom{\frac{\rd\beta_n'(\lambda_k)}{\rd\lambda_j}=}
   {} - \frac{1}{\lambda_j-\lambda_k}
      \biggl(\frac{\rd\beta_n}{\rd\lambda_k}
          + \frac{\rd\beta_{n-1}}{\rd\lambda_k}\lambda_k
          + \cdots
          + \frac{\rd\beta_1}{\rd\lambda_k}\lambda_k^{n-1} \biggl).
\end{gather*}
Obviously,
\begin{gather*}
  \frac{\rd\beta_n}{\rd\lambda_j}
  + \frac{\rd\beta_{n-1}}{\rd\lambda_j}\lambda_k
  + \cdots
  + \frac{\rd\beta_1}{\rd\lambda_j}\lambda_k^{n-1}
  = \frac{\rd\beta_n(\lambda_k)}{\rd\lambda_j},
\end{gather*}
and by \eqref{beta_n'(l_k)},
\begin{gather*}
  \frac{\rd\beta_n}{\rd\lambda_k}
  + \frac{\rd\beta_{n-1}}{\rd\lambda_k}\lambda_k
  + \cdots
  + \frac{\rd\beta_1}{\rd\lambda_k}\lambda_k^{n-1}
  = - \beta_n'(\lambda_k).
\end{gather*}
Thus the f\/irst part of \eqref{dbeta_n'(l_k)} follows.  If $j =
k$, \eqref{beta_n'(l_k)} and \eqref{d^2beta_n} imply that
\begin{gather*}
  \frac{\rd\beta_n'(\lambda_k)}{\rd\lambda_k}
  = - \frac{\rd\beta_{n-1}}{\rd\lambda_k}
    - 2\frac{\rd\beta_{n-2}}{\rd\lambda_k}\lambda_k
    - \cdots
    - (n-1)\frac{\rd\beta_1}{\rd\lambda_k}\lambda_k^{n-1}.
\end{gather*}
On the other hand, dif\/ferentiating the identity
\begin{gather*}
  \beta_n'(\lambda)
  = \beta_{n-1}(\lambda) + \beta_{n-2}(\lambda)\lambda
    + \cdots + \beta_0(\lambda)\lambda^{n-1}
\end{gather*}
(which has been used in the proof of \eqref{beta_n'(l_k)}) yields
\begin{gather*}
  \beta_n''(\lambda)
  = \beta_{n-1}'(\lambda) + \beta_{n-2}'(\lambda)\lambda
     + \cdots + \beta_1'(\lambda)\lambda^{n-1}
\\ \phantom{\beta_n''(\lambda)=}
     {} + \beta_{n-2}(\lambda) + 2\beta_{n-3}(\lambda)\lambda
     + \cdots + (n-1)\beta_0(\lambda)\lambda^{n-2}.
\end{gather*}
One can eliminate the derivatives
$\beta_{n-1}'(\lambda),\ldots,\beta_1'(\lambda)$ by the preceding
identity itself. The outcome reads
\begin{gather*}
  \beta_n''(\lambda)
  = 2\bigl(\beta_{n-2}(\lambda)
      + 2\beta_{n-3}(\lambda)\lambda
      + \cdots + (n-1)\beta_0(\lambda)\lambda^{n-2}\bigr).
\end{gather*}
Upon substituting $\lambda = \lambda_k$ and using \eqref{dbeta_n},
one f\/inds that
\begin{gather*}
  \beta_n''(\lambda_k)
  = 2\bigl(\beta_{n-2}(\lambda_k)
      + 2\beta_{n-3}(\lambda_k)\lambda_k
      + \cdots
      + (n-1)\beta_0(\lambda_k)\lambda_k^{n-2}\bigr)
\\ \phantom{\beta_n''(\lambda_k)}{}
  = - 2\biggl(
        \frac{\rd\beta_{n-1}}{\rd\lambda_k}
        + 2\frac{\rd\beta_{n-2}}{\rd\lambda_k}\lambda_k
        + \cdots
        + (n-1)\frac{\rd\beta_1}{\rd\lambda_k}\lambda_k^{n-2}
       \biggr).
\end{gather*}
The last identity and the foregoing expression of
$\rd\beta_n'(\lambda_k)/\rd\lambda_k$ lead to the second part
\linebreak of~\eqref{dbeta_n'(l_k)}.\qed

Using these lemmas, we can calculate both hand sides of
\eqref{DI_n-goal}.

Let us f\/irst consider the case of $j \not= k$.  The left hand
side of \eqref{DI_n-goal} can be calculated by the Leibniz rule
and \eqref{dbeta_n'(l_k)}. Note here that the formula
\eqref{dbeta_n'(l_k)} for $n = g$ takes such a form as
\begin{gather*}
  \frac{\rd\beta'(\lambda_k)}{\rd\lambda_j}
  = \frac{\beta'(\lambda_j)}{\lambda_j-\lambda_k}
\end{gather*}
because $\beta_g(\lambda_k) = \beta(\lambda_k) = 0$. The outcome
of this calculation reads
\begin{gather*}
  \frac{\rd}{\rd\lambda_j}
  \frac{\beta_n'(\lambda_k)}{\beta'(\lambda_k)}
  = \frac{1}{\beta'(\lambda_k)(\lambda_j-\lambda_k)}
     \frac{\rd\beta_n(\lambda_k)}{\rd\lambda_j}.
\end{gather*}
As regards the right hand side of \eqref{DI_n-goal}, we can use
\eqref{dbeta_n'(l_k)} to calculate the derivatives in the
parentheses as
\begin{gather*}
\frac{\rd\beta'(\lambda)}{\rd\lambda_k}
  \bigg|_{\lambda=\lambda_j}
  = \frac{\rd\beta'(\lambda_j)}{\rd\lambda_k}
  = \frac{\beta'(\lambda_j)}{\lambda_k-\lambda_j}, \\
\frac{\rd\beta_n'(\lambda)}{\rd\lambda_k}
  \bigg|_{\lambda=\lambda_j}
  = \frac{\rd\beta_n'(\lambda_j)}{\rd\lambda_k}
  = \frac{1}{\lambda_k-\lambda_j}
    \biggl(\frac{\rd\beta_n(\lambda_j)}{\rd\lambda_k}
      + \beta_n'(\lambda_j)\biggr).
\end{gather*}
Consequently,
\begin{gather*}
  \frac{1}{\beta'(\lambda_k)}
  \left.\left(
      \frac{\beta_n'(\lambda_j)}{\beta'(\lambda_j)}
      \frac{\rd\beta'(\lambda)}{\rd\lambda_k}
    - \frac{\rd\beta_n'(\lambda)}{\rd\lambda_k}
  \right)\right|_{\lambda=\lambda_j}
  = \frac{1}{\beta'(\lambda_k)(\lambda_j-\lambda_k)}
    \frac{\rd\beta_n(\lambda_j)}{\rd\lambda_k}.
\end{gather*}
Since \eqref{dbeta_n} implies that
\begin{gather*}
  \frac{\rd\beta_n(\lambda_j)}{\rd\lambda_k}
  = - \frac{\rd^2\beta_{n+1}}{\rd\lambda_j\rd\lambda_k}
  = \frac{\rd\beta_n(\lambda_k)}{\rd\lambda_j},
\end{gather*}
we eventually f\/ind that \eqref{DI_n-goal} holds for the case of
$j \not= k$.

The case of $j = k$ can be treated in much the same way, and turns
out to be simpler.  The left hand side of \eqref{DI_n-goal} can be
calculated as
\begin{gather*}
  \frac{\rd}{\rd\lambda_k}
  \frac{\beta_n'(\lambda_k)}{\beta'(\lambda_k)}
  = \frac{1}{2}
    \frac{\beta_n''(\lambda_k)}{\beta'(\lambda_k)}
  - \frac{1}{2}
    \frac{\beta_n'(\lambda_k)\beta''(\lambda_k)}
         {\beta'(\lambda_k)^2}.
\end{gather*}
The derivatives in the parentheses on the right hand side of
\eqref{DI_n-goal} can be expressed as
\begin{gather*}
\frac{\rd\beta'(\lambda)}{\rd\lambda_k}
 \bigg|_{\lambda=\lambda_k}
  = \frac{\rd\beta'(\lambda_k)}{\rd\lambda_k}
    - \beta''(\lambda_k)
  = - \frac{1}{2}\beta''(\lambda_k), \\
\frac{\rd\beta_n'(\lambda)}{\rd\lambda_k}
\bigg|_{\lambda=\lambda_k}
  = \frac{\rd\beta_n'(\lambda)}{\rd\lambda_k}
    - \beta_n''(\lambda_k)
  = - \frac{1}{2}\beta_n''(\lambda_k).
\end{gather*}
Thus \eqref{DI_n-goal} turns out to hold in this case, too.

We have thus conf\/irmed that \eqref{DI_n-answer} does satisfy
\eqref{DI_n-condition}. Note that \eqref{DI_n-answer} corresponds
to the correction terms
\begin{gather*}
  \Delta H_{n+1}
  = - \sum_{j=1}^g
    \frac{\mu_jR_n'(\lambda_j)}{\beta'(\lambda_j)}
\end{gather*}
for $H_{n+1}$ by the linear relation \eqref{DI_n=DH_n+...},
because $\beta_n'(\lambda)$'s and $R_n'(\lambda)$'s are linearly
related with the same coef\/f\/icients as $\beta_n(\lambda)$'s and
$R_n(\lambda)$'s. These  results can be summarized as follows.

\begin{theorem}
Equations of motion \eqref{lam-t-eq} and \eqref{mu-t-eq} can be
cast into the Hamiltonian form
\begin{gather*}
  \rd_{2n+1}\lambda_j = \{\lambda_j,K_{n+1}\},
  \qquad
  \rd_{2n+1}\mu_j = \{\mu_j,K_{n+1}\}.
\end{gather*}
The Hamiltonians are given by
\begin{gather*}
  K_{n+1}
  = \sum_{j=1}^g
    \frac{\mu_j^2 - I_0(\lambda_j)}{\beta'(\lambda_j)}
    R_n(\lambda_j)
  - \sum_{j=1}^g
    \frac{\mu_jR_n'(\lambda_j)}{\beta'(\lambda_j)}.
%\label{K_n}
\end{gather*}
\end{theorem}

\section{Examples}

We illustrate the results of the preceding section for the cases
of $g = 1,2,3$. For notational simplicity, we set $t_{2g+1} = 0$.
Consequently, $c_1(t)$ disappears from various formulas.

\subsection[$g = 1$]{$\boldsymbol{g = 1}$}

The case of $g = 1$ corresponds to the f\/irst Painlev\'e equation
itself. There is no higher f\/low (other than the excluded
exceptional time $t_3$). In this case, everything can be presented
explicitly as follows.

\begin{itemize}\itemsep=0ex
\item[1)] $\beta(\lambda)$ is linear and $\alpha(\lambda)$ does
not depend on $\lambda$:
\begin{gather*}
 \beta(\lambda) = \lambda + \beta_1,\qquad
  \alpha(\lambda) = \alpha_1, \\
 \beta_1 = R_1 = \frac{u}{2}, \qquad
  \alpha_1 = - \frac{\beta_{1,x}}{2}
          = - \frac{1}{4}u_x.
\end{gather*}
\item[2)] The Darboux coordinates $\lambda_1,\mu_1$ are given by
\begin{gather*}
  \lambda_1 = - \beta_1 = - \frac{u}{2},\qquad
  \mu_1 = \alpha(\lambda_1) = - \frac{1}{4}u_x.
\end{gather*}
\item[3)] $h(\lambda)$ is a cubic polynomial of the form
\begin{gather*}
  h(\lambda) = I_0(\lambda) + I_1, \qquad
  I_0(\lambda) = \lambda^3 + x\lambda.
\end{gather*}
\item[4)] The Hamiltonian $K_1$ is equal to $H_1 = I_1$. As a
function of the Darboux coordinates, $I_1$ can be expressed as
\begin{gather*}
  I_1 = \mu_1^2 - \lambda_1^3 - x\lambda_1,
\end{gather*}
which coincides with the well known Hamiltonian of the f\/irst
Painlev\'e equation.
\end{itemize}

\subsection[$g = 2$]{$\boldsymbol{g = 2}$}

This case corresponds to the two-dimensional `degenerate Garnier
system' studied by Kimura~\cite{Kimura89} and Shimomura
\cite{Shimomura00}. A higher f\/low with time variables~$t_3$ now
enters the game.  This variable~$t_3$ shows up in the description
of relevant quantities through
\begin{gather*}
  c_2(t) = \frac{3}{2}t_3.
\end{gather*}
For instance, the linear relations between $\beta_n$'s and $R_n$'
now read
\begin{gather*}
  \beta_1 = R_1, \;
  \beta_2 = R_2 + c_2(t).
\end{gather*}
Though slightly more complicated than the previous case, this
case, too, can be treated explicitly.

\begin{itemize}\itemsep=0pt
\item[1)] $\beta(\lambda)$ and $\alpha(\lambda)$ are quadratic and
linear, respectively:
\begin{gather*}
  \beta(\lambda)
  = \lambda^2 + \beta_1\lambda + \beta_2, \qquad
  \alpha(\lambda)
  = \alpha_1\lambda + \alpha_2.
\end{gather*}
\item[2)] The Darboux coordinates $\lambda_1,\lambda_2,
\mu_1,\mu_2$ are def\/ined as
\begin{gather*}
  \beta(\lambda)
  = (\lambda - \lambda_1)(\lambda - \lambda_2), \qquad
  \mu_1 = \alpha(\lambda_1), \qquad
  \mu_2 = \alpha(\lambda_2).
\end{gather*}
\item[3)] $h(\lambda)$ is a quintic polynomial of the form
\begin{gather*}
  h(\lambda) = I_0(\lambda) + I_1\lambda + I_2, \qquad
  I_0(\lambda) = \lambda^5 + 2c_2(t)\lambda + x\lambda.
\end{gather*}
\item[4)] We still have the simple relations $H_1 = I_1$ and $H_2
= I_2$ between $H_n$'s and $I_n$'s.  They are redef\/ined as
functions of the Darboux coordinates by the linear equations
\begin{gather*}
  I_1\lambda_1 + I_2 = \mu_1^2 - I_0(\lambda_1), \qquad
  I_2\lambda_2 + I_2 = \mu_2^2 - I_0(\lambda_2).
\end{gather*}
More explicitly,
\begin{gather*}
  I_1 = \frac{\mu_1^2 - I_0(\lambda_1)}
             {\lambda_1 - \lambda_2}
      + \frac{\mu_2^2 - I_0(\lambda_2)}
             {\lambda_2 - \lambda_2},\qquad
  I_2 = \frac{\mu_1^2 - I_0(\lambda_1)}
             {\lambda_1 - \lambda_2}\lambda_2
      - \frac{\mu_2^2 - I_0(\lambda_2)}
             {\lambda_2 - \lambda_2}\lambda_1.
\end{gather*}
The correct Hamiltonians $K_1,K_2$ are given by
\begin{gather*}
  K_1 = I_1, \qquad
  K_2 = I_2 - \frac{\mu_1}{\lambda_1 - \lambda_2}
            - \frac{\mu_2}{\lambda_2 - \lambda_1}.
\end{gather*}
\end{itemize}

\begin{remark}
Kimura and Shimomura studied this system as isomonodromic
deformations of a second order scalar ODE rather than the $2
\times 2$ matrix system. In the present setting, their scalar ODE
corresponds to the equation
\begin{gather*}
  \frac{\rd^2\psi}{\rd\lambda^2}
  + p_1(\lambda)\frac{\rd\psi}{\rd\lambda}
  + p_2(\lambda)\psi
  = 0
\end{gather*}
that can be obtained from \eqref{bspsi-l-lineq} by eliminating the
second component of $\bspsi$.\ The
coef\/f\/i\-cients~$p_1(\lambda)$ and $p_2(\lambda)$ are given by
\begin{gather*}
  p_1(\lambda)
  = \frac{\beta'(\lambda)}{\beta(\lambda)},
  \qquad
  p_2(\lambda)
  = - h(\lambda) - \alpha'(\lambda)
    + \alpha\frac{\beta'(\lambda)}{\beta(\lambda)}.
\end{gather*}
\end{remark}
\begin{remark}
Actually, Kimura and Shimomura considered two Hamiltonian forms
for their degenerate Garnier system.  One of them is def\/ined by
the aforementioned Hamiltonians $K_1$,~$K_2$. The other one is
derived therefrom by a canonical transformation, and has
polynomial Hamiltonians.
\end{remark}

\subsection[$g = 3$]{$\boldsymbol{g = 3}$}

The case of $g = 3$ is more complicated than the preceding two
cases. Here we have two higher f\/lows with time variables $t_3$,
$t_5$. They are connected with $c_2(t)$ and $c_3(t)$ as
\begin{gather*}
  c_2(t) = \frac{5}{2}t_3,\qquad
  c_3(t) = \frac{3}{2}t_5.
\end{gather*}
$h(\lambda)$ is a sextic polynomial of the form
\begin{gather*}
  h(\lambda)  = I_0(\lambda) + I_1\lambda^2  + I_2\lambda + I_3.
\end{gather*}
A new feature of this case is the structure of $I_0(\lambda)$:
\begin{gather*}
  I_0(\lambda)
  = \lambda^7 + 2c_2(t)\lambda^5
    + 2c_3(t)\lambda^4
    + (c_2(t)^2 + x)\lambda^3.
\end{gather*}
Note that the coef\/f\/icient of $\lambda^3$ is now a quadratic
polynomial of the time variables.  Of course, if we consider the
general case \eqref{h-expansion}, this is a rather common
situation; the f\/irst two cases ($g = 1$ and $g = 2$) are
exceptional.

\section{Conclusion}

We have thus elucidated the Hamiltonian structure of the PI
hierarchy for both the Lax equations and the equations of motion
in the spectral Darboux coordinates.  Though the extra
terms~$U_{n+1}'(\lambda)$ in the Lax equations give rise to extra
terms in the equations of motion for the Darboux coordinates,
these terms eventually boil down (somewhat miraculously) to the
correction terms $\Delta H_{n+1}$ in the Hamiltonian.

The correction terms $\Delta H_{n+1}$ are identif\/ied by brute
force calculations.  It is highly desirable to derive this result
in a more systematic way. As regards the Garnier system, such a
systematic explanation is implicit in the work of the Montreal
group \cite{Harnad94,HW96}, and presented (in a more general form)
by Dubrovin and Mazzocco \cite{DM03}. Let us recall its essence.

As mentioned in Introduction, the Garnier system is equivalent to
the $2 \times 2$ Schlesinger system. The $L$-matrix of the
Schlesinger system is a $2\times 2$ matrix of rational functions
of~the~form
\begin{gather*}
  V(\lambda) = \sum_{j=1}^N \frac{A_j}{\lambda - t_j}.
\end{gather*}
The matrix $A_j$ takes values in a two-dimensional coadjoint orbit
of $\mathrm{sl}(2,\CC)$. This orbit, as a~symplectic leaf, carries
special Darboux coordinates $\xi_j$, $\eta_j$. $A_j$ can be
thereby written as
\begin{gather*}
  A_j = \frac{1}{2}
        \left(\begin{matrix}
        \xi_j\eta_j & \xi_j^2 \\
        - \eta_j^2 + \theta_j^2\xi_j^{-2} & - \xi_j\eta_j
        \end{matrix}\right),
\end{gather*}
where $\theta_j$ is a constant that determines the orbit, and may
be interpreted as a monodromy exponent at the regular singular
point at $\lambda = t_j$.   The Lax equations can be converted to
a Hamiltonian system in these Darboux coordinates $\xi_j$,
$\eta_j$. Since the spectral Darboux
coordinates~$\lambda_j$,~$\mu_j$ are connected with these Darboux
coordinates by a {\it time-dependent} canonical transformation,
the Hamiltonians in the latter coordinates have extra terms.

Unfortunately, this beautiful explanation of extra terms does not
literally apply to the present setting.  The relevant Lie algebra
for this case is not $\mathrm{sl}(2,\CC)$ but its loop algebra
$\mathrm{sl}(2,\CC)[\lambda,\lambda^{-1}]$, coadjoint orbits of
which are more complicated.

A similar idea, however, can be found in the recent work of
Mazzocco and Mo \cite{MM06} on an isomonodromic hierarchy related
to the second Painlev\'e equation. They start from the Lie-Poisson
structure of a loop algebra, and convert the Hamiltonian structure
on a coadjoint orbit to a Hamiltonian system in Darboux
coordinates by a time-dependent canonical transformation.  It will
be interesting to reconsider the present setting from that point
of view.

Another remarkable aspect of the work of Mazzocco and Mo is that
they present another set of Darboux coordinates alongside the
spectral Darboux coordinates.  Unlike the spectral Darboux
coordinates, these coordinates are rational functions of the
dynamical variables in the Lax equations;  this is a desirable
property in view of the Painlev\'e property of the system.

Actually, borrowing their idea, we can f\/ind a similar set of
Darboux coordinates $\mathcal{Q}_n$, $\mathcal{P}_n$ for the PI
hierarchy as
\begin{gather*}
  \mathcal{Q}_n = \beta_{g+1-n}, \qquad
  \mathcal{P}_n = \sum_{k=1}^g
           \frac{\rd p_n}{\rd\beta_k}
           \frac{\alpha_k}{n},
\end{gather*}
where $p_k$ stands for the $k$-th power sum
\begin{gather*}
  p_k = \sum\limits_{j=1}^g \lambda_j^k.
\end{gather*}
(Note that $p_k$'s and $\beta_j$'s are related by the generating
functional relation
\begin{gather*}
  \sum_{k=1}^\infty \frac{p_k}{k\lambda^k}
  = - \log \frac{\beta(\lambda)}{\lambda^g},
\end{gather*}
so that $p_k$'s may be thought of as polynomial functions of
$\beta_j$'s.) One can prove, in the same way as the case of
Mazzocco and Mo, that $\mathcal{Q}_n$, $\mathcal{P}_n$ do satisfy
the canonical Poisson relations. In fact, they turn out to satisfy
the stronger relation
\begin{gather*}
    \sum_{j=1}^g d\lambda_j \wedge d\mu_j
  = \sum_{n=1}^g d\mathcal{Q}_n \wedge d\mathcal{P}_n,
\end{gather*}
which implies that $\mathcal{Q}_n$, $\mathcal{P}_n$ are connected
with the spectral Darboux coordinates $\lambda_j$, $\mu_j$ by
a~{\it time-independent} canonical transformation.  In particular,
the canonical transformation yields no correction term to the
transformed Hamiltonians. Namely, $K_{n+1}$'s persist to be
correct Hamiltonians in the new coordinates $\mathcal{Q}_n$,
$\mathcal{P}_n$ as well.  Moreover, employing the Lagrange
interpolation formula, one can see that $K_{n+1}$'s are
polynomials in these coordinates (and the time variables).  We
thus obtain a generalization of Kimura's polynomial Hamiltonians
for the degenerate Garnier systems \cite{Kimura89}.

\subsection*{Acknowledgments}

I would like to thank G. Falqui, E. Inoue and M. Mazzocco for
valuable comments and discussions. This research was partially
supported by Grant-in-Aid for Scientif\/ic Research No. 16340040
from the Japan Society for the Promotion of Science.

\pdfbookmark[1]{References}{ref}
\LastPageEnding

\end{document}